\documentclass[aoas]{imsart}

\RequirePackage{amsthm,amsmath,amsfonts,amssymb}
\RequirePackage[authoryear]{natbib}
\RequirePackage[colorlinks,citecolor=blue,urlcolor=blue]{hyperref}
\RequirePackage{graphicx}

\usepackage{amsmath}
\usepackage{lipsum}
\usepackage{amsfonts}
\usepackage{amssymb}
\usepackage{amsthm}
\usepackage{bbm}
\usepackage{booktabs}

\startlocaldefs
\theoremstyle{plain}

\theoremstyle{remark}

\newcommand{\vect}[1]{\textbf{#1}}
\newcommand{\matr}[1]{\textit{\textbf{#1}}}

\endlocaldefs

\begin{document}

\begin{frontmatter}
\title{Multi-model Ensemble Analysis with Neural Network Gaussian Processes}
\runtitle{Climate Model Integration}

\begin{aug}
\author[A]{\fnms{Trevor} \snm{Harris}\ead[label=e1,mark]{tharris@tamu.edu}},
\author[B]{\fnms{Bo} \snm{Li}\ead[label=e2]{libo@illinois.edu}}
\and
\author[C]{\fnms{Ryan} \snm{Sriver}\ead[label=e3]{rsriver@illinois.edu}}
\address[A]{Department of Statistics,
Texas A\&M University,
\printead{e1}}

\address[B]{Department of Statistics,
University of Illinois at Urbana Champaign,
\printead{e2}}

\address[C]{Department of Atmospheric Sciences,
University of Illinois at Urbana Champaign,
\printead{e3}}

\end{aug}

\begin{abstract}
Multi-model ensemble analysis integrates information from multiple climate models into a unified projection. However, existing integration approaches based on model averaging can dilute fine-scale spatial information and incur bias from rescaling low-resolution climate models. We propose a statistical approach, called NN-GPR, using Gaussian process regression (GPR) with an infinitely wide deep neural network based covariance function. NN-GPR requires no assumptions about the relationships between climate models, no interpolation to a common grid, and automatically downscales as part of its prediction algorithm. Model experiments show that NN-GPR can be highly skillful at surface temperature and precipitation forecasting by preserving geospatial signals at multiple scales and capturing inter-annual variability. Our projections particularly show improved accuracy and uncertainty quantification skill in regions of high variability, which allows us to cheaply assess tail behavior at a 0.44$^\circ$/50 km spatial resolution without a regional climate model (RCM). Evaluations on reanalysis data and SSP2-4.5 forced climate models show that NN-GPR produces similar, overall climatologies to the model ensemble while better capturing fine scale spatial patterns. Finally, we compare NN-GPR's regional predictions against two RCMs and show that NN-GPR can rival the performance of RCMs using only global model data as input.

\end{abstract}


\end{frontmatter}

\section{Introduction} \label{sec:introduction}

\newdimen\stringwidth
\setbox0=\hbox{$\emptyset$}
\stringwidth=\wd0

Climate models are central to modern climate science and are the primary tool for projecting future climate states \citep{flato2014evaluation}. However, constructing climate models is an international effort, with various modeling centers developing distinct models semi-independently. This situation has led to many plausible, though disagreeing, models all representing the same earth system \citep{knutti2010challenges, flato2014evaluation}. Rather than select a ``best'' climate model, climate scientists incorporate many models into multi-model ensembles that are combined or integrated into a consensus estimate. Model integrations often show improved reconstruction skill compared to the individual models \citep{lambert2001cmip1, gleckler2008performance, knutti2010challenges}. Ensembles also allow climate scientists to quantify uncertainty through inter-model variability. For example, initial condition ensembles sample internal variability, perturbed physics ensembles sample uncertainties in parameters, and multi-model ensembles such as CMIP sample forcing uncertainties, structural model differences, and initial conditions \citep{sansom2017constraining}.

Properly integrating multi-model ensembles into a consensus estimate has been the topic of much discussion \citep{tebaldi2007use}. The most common and convenient approach is to average all models into a pointwise ensemble mean \citep{flato2014evaluation} and use the pointwise inter-model variability as the projection uncertainty. Averaging is done either democratically, whereby each model receives the same weight or with weights based on model skill, reliability, and inter-model dependence \citep{giorgi2002calculation, giorgi2003probability, abramowitz2019esd}. This type of model integration is known as ensemble averaging or weighted ensemble averaging.

\begin{figure}
	\centering
	\includegraphics[width=0.9\textwidth]{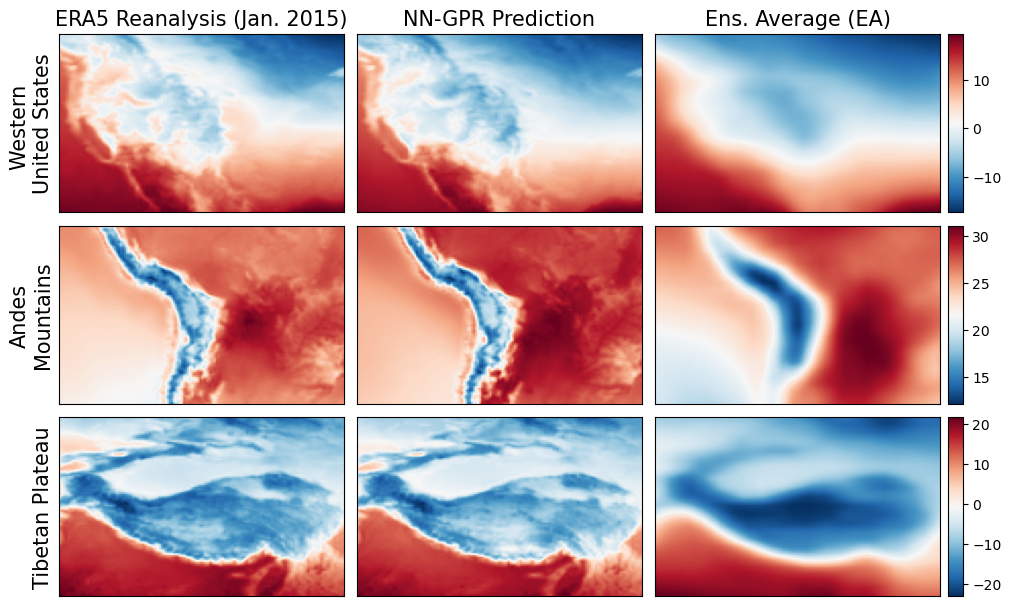}
    \caption{ERA5 reanalysis field (Section \ref{sec:data}) for January 2015 (left), our proposed method NN-GPR's (Section \ref{sec:model}) prediction (center) and an ensemble average of the 16 climate models (right) described in Section \ref{sec:data}. The ensemble average (EA) shows heavy blurring due to resizing, re-interpolating, and averaging low resolution climate model output on a high resolution grid. Our approach (NN-GPR) uses the same climate model ensemble to make detailed predictions and requires no further post-processing to downscale. Temperature scales in $^\circ$C.}
    \label{fig:prediction_viz}
\end{figure}

A fundamental challenge with ensemble averaging is how to gain the benefits of a model average without losing individual model skill, i.e., while retaining inter-annual variability and spatial information. In the extreme case, simple model averaging may cause severe blurring that erodes nearly all spatial signals (Figure \ref{fig:prediction_viz}). This drawback has spurred many alternative approaches based on reliability weighting, Bayesian hierarchical models, regression, and machine learning that all use observational data to improve model integration.

Bayesian methods assume a statistical model for the actual climate and use the climate models and observations to learn the parameters of the statistical model \citep{tebaldi2004regional, smith2009Bayesian, bhat2011climate, rougier2013second, sansom2017constraining, bowman2018hierarchical}. Bayesian approaches differ significantly from ensemble averaging methods in incorporating model variability to describe the actual climate. Instead of linearly combining models or picking a subset to combine linearly, Bayesian models quantify different sources of variability (e.g., model uncertainty, model inadequacy, and natural variability) and learn posterior estimates from the climate model ensembles. As a result, the posterior estimates can rigorously quantify prediction uncertainty and make distributional forecasts. Bayesian methods can also learn more complex, emergent relationships than simple linear combinations \cite{chandler2013exploiting, sansom2017constraining}.

Regression \citep{raisanen2010weighting, bracegirdle2012higher} and machine learning methods \citep{ghafarianzadeh2013climate} learn a mapping from the climate model ensemble, or the ensemble mean, to the target climate process. These prediction oriented methods often show superior predictive skill compared to ensemble averaging \citep{greene2006probabilistic}. However, they can be computationally expensive and may fail to generalize to future climate scenarios if the distribution of the climate models changes significantly from the training period. Additionally, linear regression strongly assumes that the actual climate system is a simple linear combination of model output.

Instead, we propose a nonparametric approach and use Gaussian process regression (GPR) \citep{rasmussen2003gaussian} to combine climate models. We use the recently developed deep neural network kernel functions \citep{lee2017deep, garriga2018deep} to define a GPR model, called NN-GPR, from climate model ensembles to reanalysis fields that exactly reproduces the predictions of an infinitely wide deep neural network. GPR combines the strengths of Bayesian methods and prediction methods into a fundamentally new approach for highly flexible model integration.

Our approach requires no assumptions about the relationships between models, no interpolation to a common grid, and is computationally efficient at our sample sizes. Additionally, NN-GPR has only three parameters yet still allows for non-linear and dynamically evolving predictions, which we show can outperform existing averaging and regression approaches. We show that our approach can preserve geospatial characteristics beyond the capabilities of previous methods and thus act as a simultaneous forecasting and pattern scaling technique (Figure \ref{fig:prediction_viz}) comparable to regional climate models (RCMs). By using low resolution model output to capture high resolution patterns, we partially mitigate the high computational cost of RCM ensembling while still capturing tail behavior through the posterior predictive distributions.

\section{Data} \label{sec:data}

We consider three publicly available data sets to develop and evaluate our integration method. These include a climate model ensemble observed under historical and future forcing scenarios, a reanalysis data product, and two regional climate models.

In Section \ref{sec:experiments} we first use the climate model ensemble by jackknifing out GCM simulation to serve as surrogate observations in order to validate our approach on explicit, physically based future simulations. This cannot be done with historical reanalysis data, which is only split into historical training and validation periods. In Section \ref{sec:application}, we then train our model to predict reanalysis data from the climate model ensemble and apply it to future projections under SSP2-4.5 to qualitatively compare our approach against several baseline methods. We further compare our predictions, subset to North America, against regional climate models in Section \ref{sec:rcm_comparisons}.

\subsection{Climate models} \label{sec:data_models}

We consider a small 16-member ensemble of monthly average 2-meter surface temperature (T2M) fields and average total precipitation (PR) fields. Our ensemble includes ACCESS-CM2, BCC-CSM2-MR, CMCC-CM2-SR5, CanESM5-CanOE, two ensemble runs of CanESM5, FIO-ESM-2-0, GFDL-ESM4, INM-CM5-0, two ensemble runs of IPSL-CM6A-LR, KACE-1-0-G, MCM-UA-1-0, and two ensemble runs of MIROC-ES2L \citep{eyring2016overview}. For each model, we have gridded T2M and PR monthly averages. Each model is observed on either a 100, 250, or 500km grid, depending on the model. All models are run under ensemble setting r1p1f1 in the Coupled Model Intercomparison Project (CMIP6) \citep{eyring2016overview}.

From each climate model, we have gridded output under the historical forcing scenario from January 1950 through December 2015. We also have gridded output from each model run under socioeconomic scenario 2 with representative concentration pathway 4.5, hereafter SSP2-4.5, from January 2016 through December 2099. To align with the reanalysis data availability (1950-2021), we will, unless otherwise stated, concatenate the historical output (Jan. 1950 - Dec. 2014) with the SSP2-4.5 output (Jan. 2015 - Dec. 2020) to form training data. Thus the training inputs will include mostly historical climate model output and some SSP2-4.5 output. The remainder of the SSP2-4.5 output from (Jan. 2021 - Dec. 2099) will be used as testing data.
SSP2-4.5 was chosen as a plausible middle-of-the-road forcing scenario that “reflects an extension of the historical experience” \citep{fricko2017marker}. 

\subsection{Reanalysis}  \label{sec:data_reanalysis}

We use gridded reanalysis data to calibrate our climate model integration as a surrogate for proper observational data. Reanalysis is an optimal combination of observations and results from a weather forecasting model that provides a global gridded representation of the Earth’s weather/climate variables on sub-daily timescales going back to 1950. The weather model helps fill in the gaps where there are no observations, and the observations are assimilated into the reanalysis to ensure the model remains close to reality. Reanalysis is heavily based on observational data but uses models for assimilation, so it is not considered entirely observational but quasi-observational.

Reanalysis data come from the ERA5 reanalysis product \citep{hersbach2020era5}, which used the Integrated Forecasting System (IFS) Cy41r2, and spans January 1950 through December 2021. For integrating T2M climate model output, we calibrate with reanalysis 2-Meter Surface Temperature fields (T2M) on single pressure levels. For integrating PR climate model output, we use Total Precipitation fields (PR) on single pressure levels. In both cases, the reanalysis fields are gridded to a 30km grid which is substantially finer than any of the climate models. Because the target reanalysis fields are of much high resolution than the climate model fields, this leads to a statistical downscaling effect that we explore in Section \ref{sec:rcm_comparisons}. Note, however, that the climate model ensemble members that compose the input training data set for NN-GPR are not regridded to the reanalysis grid.

We will treat ERA5 reanalysis fields as ground truth to calibrate our model integration method. However, other data products, such as NCEP \citep{kalnay1996ncep}, could also have been used to represent observational data. Different reanalysis products do not necessarily agree, so calibration with an alternative product could change our empirical findings (Section \ref{sec:application}). However, the methodology would not change.

\subsection{Regional climate models}

In Section \ref{sec:application} we investigate the downscaling skill of our model and compare it with regional climate model (RCM) output. We focus on North America, so we use the output from NA-CORDEX simulations \citep{mearns2017cordex} from 1950-2021. Specifically, we use CanESM2 projections downscaled via CRCM5-OUR and CanRCM4 onto a 25km grid cropped to match the output of our method. The two RCM models are also averaged to produce a single RCM estimate.

\section{Background} \label{sec:background}
Since our proposed approach is a Gaussian process regression (GPR) model \citep{rasmussen2003gaussian} empowered with deep neural network kernels (NNGP) \citep{lee2017deep, garriga2018deep}, we briefly review GPR and NNGP below. 

\subsection{Gaussian process regression}
Let $D = \{(x_i, y_i)\}_{i=1}^n$ represent a dataset of $n$ paired observations $(x_i, y_i) \in \mathbbm{R}^q \times \mathbbm{R}^d$. In this section, we assume $d =1$ for an easy exposition of the concept, but the GPR can be easily extended to $d > 1$ as in our model in Section \ref{sec:model}, where $x_i$ is a climate model ensemble and $y_i$ is a reanalysis field. 

In Gaussian process regression, we model an unknown, univariate regression function $f: X \in \mathbbm{R}^q \mapsto Y \in \mathbbm{R}$ as a Gaussian process ($\mathcal{GP}$) with mean function $m: \mathbbm{R}^q \mapsto \mathbbm{R}$ and covariance kernel $K: \mathbbm{R}^q \times \mathbbm{R}^q \mapsto \mathbbm{R}$. A Gaussian process with mean $m$ and covariance $K$ can be defined on random functions $f$ by assuming any finite set of evaluations of $f$, say $\vect{f} = (f(x_1),\ldots,f(x_n))^T$, where $^T$ represent transpose, is multivariate normal with mean vector $\vect{m}$ and covariance matrix $\matr{K}$. That is, if for any collection $x_1,...,x_n$ we have 
	\begin{equation*}
		\vect{f} \sim MVN(\vect{m}, \matr{K}),
	\end{equation*}
where $\vect{m} = (m(x_1),...,m(x_n))$ is the $n$ dimensional vector such that $m(x_i)=E(f(x_i))$ and $\matr{K}$ is a $n \times n$ dimensional symmetric, positive definite matrix whose entries are determined by the covariance function $K$, then the random function $f$ follows a Gaussian process with mean function $m$ and covariance function $K$. Following \citet{van2011information}, we denote this as
	\begin{equation*}
		f \sim \mathcal{GP}(m, K).
	\end{equation*}
To model the response variables $y_1,...,y_n$ as a function of $x_1,...,x_n$, GPR typically assumes the following additive model with Gaussian error
\begin{equation} \label{eqn:gpr}
	\begin{aligned}
    y_i &= f(x_i) + \epsilon_i, \\
    f &\sim \mathcal{GP}(m, K), \\
    \epsilon_i & \sim \hbox{ iid }N(0, \sigma^2),
	\end{aligned}
\end{equation}
where iid means independent and identically distributed, $m$ is a mean function and $K$ is a covariance function.

Gaussian process regression is a type of nonparametric regression because we place a prior on the function $f$, rather than on parameters defining $f$, e.g. assuming $f$ is linear $f(x) = \beta x$ and modeling $\beta$. With the assumption that $\epsilon$ is Gaussian noise, the posterior of $f$ is also a Gaussian process by conjugacy of the multivariate normal distribution \citep{rasmussen2003gaussian}. Moreover, we can analytically compute the posterior predictive distribution of $f(x_0)$, based on a new observation $x_0$ and the training data $D$, as
\begin{equation} \label{eqn:gpr_predictions}
	\begin{aligned}
    P(f(x_0) \mid x_0, D) &= N(\mu_0, K_0), \\
	\mu_0 &= m(x_0) + \vect{k}(x_0)(\matr{K} + \sigma^2\matr{I}_n)^{-1} \vect{y}^T, \\
    K_0 &= K(x_0, x_0) - \vect{k}(x_0)(\matr{K} + \sigma^2 \matr{I}_n)^{-1} \vect{k}^T(x_0),
	\end{aligned}
\end{equation}
where $\vect{y} = (y_1, \ldots, y_n)$, $\vect{k}(x_0) = (K(x_0, x_1),...,K(x_0, x_n))$, $\matr{I}_n$ is an $n \times n$ identity matrix, and $\matr{K}$ is an $n \times n$ covariance matrix with entries determined by the covariance function $K$. 
Thus, given the prior mean and covariance functions, and historical data, we can immediately compute the posterior predictive distribution of $f(x_0)$ at any test point $x_0$. Consequently, the posterior prediction distribution of $y_0$ is $P(y_0 \mid x_0, D) = N(\mu_0, K_0 + \sigma^2)$. 

\subsection{Deep neural network kernels} \label{sec:background:nngp}
The mean function, $m$, can often be set to a constant, such as 0, for a stationary process or to some (non)linear trend over $\mathbb{R}^q$. The covariance function, $K$, is typically more difficult to specify, particularly for complex objects such as ensembles of climate fields. Some recent works have developed complex covariance functions through low rank approximations \citep{katzfuss2017multi} or dimension augmentation \citep{bornn2012modeling,shand2017modeling}.
While these approaches provide general covariance functions for GPR, they may be difficult to justify or compute for our particular problem of mapping ensembles of climate models to reanalysis fields.

In \cite{lee2017deep}, the authors showed that deep Bayesian neural networks \citep{mackay1992practical, neal2012Bayesian, wilson2020Bayesian} define valid, typically non-stationary, covariance functions. They showed that untrained, infinitely wide, Bayesian neural networks (BNN) are equivalent to Gaussian processes with a recursively defined covariance function. This family of Gaussian processes is termed Neural Network Gaussian Processes (NNGP). 

We review the general form of the NNGP covariance function and present a special case based on Rectified Linear Unit (ReLU) activations that we use in our study. ReLU is a piecewise linear function that will output its input when the input is positive and outputs zero otherwise. It is the default activation function for many types of neural networks \citep{ramachandran2017searching} and has a closed form covariance function. Let $\Phi_A$ denote an untrained, fully connected deep neural network with architecture $A$, and mapping from $\mathbb{R}^q$ to $\mathbb{R}^d$. Fully connected neural networks are a type of artificial neural network where the architecture connects all the neurons in one layer to those in the next layer \citep{goodfellow2016deep}. We denote the $r$'th output of the $l$'th layer of $\Phi_A$ given an input $x_i \in \mathbbm{R}^q$ as 
\begin{equation} \label{eqn:bnn}
	Z_r^l(x_i) = b_r^l + \sum_{k = 1}^{N_l}W_{rk}^l \phi(Z_k^{l-1}(x_i)),\\
\end{equation}
and $$Z_r^0(x_i) = b_r^0 + \sum_{k=1}^{q} W_{rk}^0 x_i(k),$$ 
where $x_i(k)$ is the $k$th element of $x_i$, $\phi$ is an activation function, such as ReLU or Tanh \citep{goodfellow2016deep}, and $N_l$ is the layer width. We further assume that the weights ($W_{rk}^l$) and biases ($b_r^l$) of $\Phi_A$ follow symmetric, independent, zero mean prior distributions with prior variances $Var(W_{rk}) = \sigma^2_w/N_l$ and $Var(b_r) = \sigma^2_b$.

By letting $N_l \rightarrow \infty$, the distribution of $Z_r^l(x_i)$ will converge to a Gaussian distribution with mean zero and a finite variance. Furthermore, $Z_r^l(x_1), \ldots, Z_r^l(x_n)$ will jointly have a Gaussian process distribution with mean zero and covariance between any $Z_r^l(x_i)$ and $Z_r^l(x_j)$ as     
\begin{equation} \label{eqn:nngp_cov}
K^l(x_i, x_j) = \sigma_b^2 + \sigma_w^2 \mathbb{E}_{Z_r^{l-1}}[\phi(Z_r^{l-1}(x_i))\phi(Z_r^{l-1}(x_j))],
\end{equation}
by the central limit theorem \citep{lee2017deep,garriga2018deep}. Here $\mathbb{E}_X[\cdot]$ denotes the expected value operator with respect to $X$. \citet{lee2017deep} further derived that the expectation in (\ref{eqn:nngp_cov}) can be written into
\begin{equation}\label{eqn:nngp_F}
	\mathbb{E}_{Z_r^{l}}[\phi(Z_r^l(x_i))\phi(Z_r^l(x_j))]= F_\phi(K^l(x_i, x_i), K^l(x_i, x_j), K^l(x_j, x_j)),
\end{equation}
for a deterministic function $F_\phi$.

Assuming $\Phi_A$ has $L$ layers, we can combine (\ref{eqn:nngp_cov}) and (\ref{eqn:nngp_F}) to write the output covariance $K^L(x_i, x_j)$ as a sequence of non-linear transformations
\begin{equation} \label{eqn:NNGP_fullcov}
\begin{aligned}
	K^L(x_i, x_j) &= \sigma_b^2  + \sigma_w^2 F_\phi(K^{L-1}(x_i, x_i), K^{L-1}(x_i, x_j), K^{L-1}(x_j, x_j)), \\
	&\vdots \\
	K^1(x_i, x_j) &= \sigma_b^2 + \sigma_w^2 F_\phi(K^0(x_i, x_i), K^0(x_i, x_j), K^0(x_j, x_j)),\\
	K^0(x_i, x_j) &= \sigma_b^2 + \sigma_w^2 (x_i^T x_j/q),
\end{aligned}
\end{equation}
which reveals a recursive structure. 
In many cases, $F_\phi(K^l(x_i, x_i), K^l(x_i, x_j), K^l(x_j, x_j))$ can be computed analytically for each $l \in 1,...,L$. For example, if $\phi$ is a ReLU activation, then
\begin{align*}
   & F_\phi(K^l(x_i, x_i), K^l(x_i, x_j), K^l(x_j, x_j))  \\ 
   &\quad = \frac{1}{2\pi} \sqrt{K^l(x_i, x_i)K^l(x_j, x_j)} \left( \sin(\theta^l_{x_i, x_j}) + (\theta^l_{x_i, x_j} - \pi) \cos(\theta^l_{x_i, x_j}) \right ),
\end{align*}
where
\begin{align*}
	\theta^l_{x_i, x_j} &= \arccos \left( \frac{K^l(x_i, x_j)}{\sqrt{K^l(x_i, x_i)K^l(x_j, x_j)}} \right).
\end{align*}

With $K^L$, we can define a GPR that exactly mimics the predictions of an infinitely wide neural network $\Phi_A$. That is, if we had the posterior distribution of all weights and biases to integrate their randomness out in the $\Phi_A$ prediction, then the neural network results would be identical to the GPR with the $K^L$ kernel.  
For finite-width networks, GPR will approximate the predictions of $\Phi_A$

\section{Model} \label{sec:model}
In the context of our data, we denote the training dataset as the sequence $D = \{(x_t, y_t)\}_{t=1}^T$, where each $x_t$ represents an \textit{ensemble} of $m$ gridded climate model fields all observed at time $t$, and each $y_t$ represents the gridded reanalysis field at time $t$. We represent each $x_t$ as a vector by vectorizing each of the $m$ fields at time $t$ and concatenating the resulting vectors into one vector. We define $q=\sum_{i=1}^m q_i$, where $q_i$ is the number of grid points in the $i$th model, as the dimension of the ensemble. We also represent each $y_t$ as a vector of length $d$, the number of grid points in the reanalysis field.

Because $d$ is very large in our application, it is computationally infeasible to define a multi-output Gaussian process over the vector $y_t$, i.e. to define an $f: \mathbbm{R}^q \mapsto \mathbbm{R}^d$ directly.
Instead, we define $d$ component functions $f_s : \mathbbm{R}^q \mapsto \mathbbm{R}$, where $s \in \{s_1,...,s_d \}$ denotes a spatial location in $y_t$, independently as
\begin{equation} \label{eqn:gpr_model}
	\begin{aligned}
    y_t(s) &= \bar{\vect{x}}_t \boldsymbol{\beta} + f_s(x_t) + \epsilon_{s, t}, \\
    f_s & \sim \mathcal{GP}\left(0, K_{\theta} \right), \\
    \epsilon_{s, t} & \sim \hbox{ iid } N(0, \sigma^2)
	\end{aligned}
\end{equation}
where $y_t(s)$ is $y_t$ at location $s$. The first term, $\bar{\vect{x}}_t \boldsymbol{\beta}$, is intended to keep the prediction of $y_t(s)$ calibrated towards the climate model ensemble mean at the same time. Here $\bar{\vect{x}}_t = (\bar{x}_{t, 1},...,\bar{x}_{t, m})$ are the spatial means for each of the $m$ climate model outputs comprising $x_t$.The second term, $f_s(x_t)$, describes how we transform the climate model ensemble $x_t$ into a single point on the reanalysis field $y_t(s)$ after accounting for the trend.
Finally, $\epsilon_{s, t}$ represents the \textit{i.i.d} white noise residual over time and space after accounting for the trend and the Gaussian process prediction. The residual variance, $\sigma^2$, is constant based on the assumption that $f_1,...,f_d$ captures all heterogeneous spatial variability in $y_t$.

We place a Gaussian process prior on each $f_s$, which has mean zero and an NNGP based covariance kernel $K_\theta$, where $\theta=\{\sigma^2_w, \sigma^2_b\}$. We define $K_\theta(x_i, x_j) = K^L(x_i, x_j)$, for each pair of climate model ensembles $x_i, x_j$, where $K^L(x_i, x_j)$ is defined in Equation (\ref{eqn:NNGP_fullcov}) assuming ReLU activations. We choose an NNGP kernel over simpler parametric kernels because it is a powerful and flexible tool for representing complex dependence patterns in high dimensional space. This is particularly relevant in our application because the evaluations $f(x_1),...,f(x_T)$ are based on the very high dimensional climate model ensembles $x_1,...,x_T$, which have an unknown correlation structure. The NNGP kernel ensures that our GPR inherits the great power of deep neural networks, while enjoying a significantly reduced computational cost as long as the $T\times T$ covariance matrix inversion is feasible.  

Compared to standard deep learning, the NNGP can be trained with substantially fewer observations due to the structure provided by Gaussian process regression and the fact that we only need to train three parameters, rather than several million. We crucially assumed that $K_{\theta}$ is the same for all spatial locations $s \in 1,\dots, d$, i.e. that the parameters are shared so we only need to estimate a single parameter set. Therefore, our approach is ideally suited for learning complex functions from high-dimensional vectors (climate field output) to high-dimensional vectors (reanalysis data) with limited training data \citep{arora2019harnessing}.

\textit{Spatial and temporal correlation} ---
Although our model (\ref{eqn:gpr_model}) models the data at each location separately, the covariance kernel $K_{\theta}$ in the $\mathcal{GP}$ prior is identical for all spatial locations $s \in 1,\dots, d$. Thus, spatial dependency patterns shared across all modeled climate fields and reanalysis data are partially reckoned in the optimization of $\theta$, since $\theta$ is shared among all locations. This is an explicit and also a different way of accounting for spatial dependency in data analysis, compared to the traditional approaches that often use neighborhood information when making inferences for spatial data. For temporal correlations, the proposed model could be extended to include lagged ensemble inputs, e.g. mapping $\text{concat}(x_{t-1}, x_t)$ to $y_t$. However, in a simulation study (Appendix B.1), we found no direct benefit of including lagged ensembles.

\textit{Parameters} ---
Our model has $m+3$ unknown parameters, $\boldsymbol\beta, \sigma^2_w, \sigma^2_b$ and $\sigma^2$. 
Rather than placing hyperpriors on the unknown parameters $\boldsymbol\beta, \sigma^2_w, \sigma^2_b$ and $\sigma^2$, we estimate them using the maximum marginal likelihood approach for Gaussian process regression \citep{rasmussen2003gaussian}. We discuss our parameter estimation strategy in Section A of the appendix.

In addition to these parameters, there are additional ``hyperparameters'' that determine the form of $K_{\theta}$, including the choice of activation function, the layer type / inductive biases, and the depth of the underlying network that cannot be optimized. These parameters define the architecture $A$ of the underlying neural network $\Phi_A$, which ultimately defines  $K_{\theta}$. For simplicity, we employ fully connected layers with ReLU activations in our NNGP since this yields an analytical form for $K_{\theta}$. We found this simple architecture outperformed existing methods in our experiments, and hence more complex layers, such as convolutional layers \citep{goodfellow2016deep}, were not necessary. For the network depth $L$, we found that predictive performance is relatively insensitive to $L$ for $1 \leq L \leq 20$ (Section B.2.). For all experiments (Section \ref{sec:experiments}) and reanalysis results (Section \ref{sec:application}), we use $L = 10$ as a compromise between accuracy (over short and long term prediction) and runtime.

\textit{Prediction} ---
Given a new multi-model ensemble, say $x_F$ for a future time $F > T$, we can integrate $x_F$ into a probabilistic forecast by computing the posterior prediction distribution of the unobserved reanalysis field $y_F$. Let $y_F(s)$ denote $y_F$ at the location $s$, then 

\begin{equation}\label{eqn:model-pred}
	\begin{aligned}
    P(y_F(s) \mid x_F, D) &= N(\mu_F(s), K_F + \sigma^2), \\
	\mu_F(s) &= \bar{\vect{x}}_F \boldsymbol{\beta} + \vect{k}_{\theta}(x_F)\left(\matr{K}_{\theta} + \sigma^2\matr{I}_T\right)^{-1} \left[\vect{y}(s) - \bar{\vect{x}}_F \boldsymbol{\beta}  \right]^T, \\
    K_F &= K_{\theta}(x_F, x_F) - \vect{k}_{\theta}(x_F)\left(\matr{K}_{\theta} + \sigma^2 \matr{I}_T\right)^{-1} \vect{k}^T_{\theta}(x_F).
	\end{aligned}
\end{equation}
where $\matr{K}_{\theta}$ is a $T\times T$ matrix whose entries are defined by $K_{\theta}$ (Equation \ref{eqn:NNGP_fullcov}), $\vect{y}(s) = (y_1(s),...,y_n(s))$, $\vect{k}_{\theta}(x_F) = (K_{\theta}(x_F, x_1),...,K_{\theta}(x_F, x_n))$, and $\bar{\vect{x}}_F = (\bar{x}_{F, 1},...,\bar{x}_{F, m})$ are the spatial means for each of the $m$ climate model outputs comprising $x_F$.
We compute $P(y_F(s) \mid x_F, D)$ for all $s \in s_1,...,s_d$  so that we can predict the field $y_F$ pointwise as $\hat{y}_F(s) = \mu_F(s)$ and quantify uncertainty pointwise with $K_F + \sigma^2$. Thus our model has a spatially varying posterior predictive mean, but a spatially constant posterior predictive variance.

\section{Numerical experiments in model prediction} \label{sec:experiments}

We compare our model, NN-GPR, against baseline methods: ensemble averaging (EA), weighted ensemble averaging (WEA) \citep{knutti2017climate}, ensemble regression (LM) \citep{bracegirdle2012higher}, a convolutional neural network (CNN) \citep{goodfellow2016deep}, Gaussian process regression using exponential (GPEX) and squared exponential (GPSE) kernels, and the delta method (DELT). We exclude hierarchical Bayesian methods \citep{sansom2017constraining} from comparison due to their complexity, computational cost, and lack of software implementation for our problem.

We use 16 ``perfect model'' tests to compare integration methods. In each test, one model is held out from the ensemble to become the target $y_t$ and the remaining 15 models become $x_t$ \citep{knutti2017climate}. The key advantage of this approach is that explicit, physically based future projections are available for validation, unlike with reanalysis data. We use our method and the seven baseline approaches to predict $y_t$ with $x_t$ and average prediction and uncertainty quantification scores across all 16 experiments. We train each model using monthly averaged fields from 1950 through 2021, and test models on monthly averaged fields from 2022-2100 as described in Section \ref{sec:data_models}.

\subsection{Existing approaches}

The first type of method we consider is the model averaging approach, which includes EA, WEA, and LM. To construct the ensemble average, we resize each model to the target field's resolution and pointwise average the resized fields. The weighted ensemble average is the same, except that each model is weighted by its correlation with the target field as described in \cite{knutti2017climate}. For the ensemble regression, we again resize each model to the target resolution, then compute a linear regression between the resized model ensemble and the target field at each spatial location. We do not include an intercept.

We then consider the pure predictive approach, which includes the CNN, GPEX, and GPSE. These methods take the climate model ensemble as input and predict the corresponding reanalysis field. For the CNN, we use a small fully convolutional architecture ($128 \times 64 \times 34 \times 1$ filters) trained with the Adam optimizer \citep{kingma2014adam}. Again each model is resized to the target resolution first, so no resizing occurs within the deep neural network. For the Gaussian process regressions, GPEX and GPSE, we reuse the model in Equation \ref{eqn:gpr_model} except for replacing the NNGP kernel with either a squared exponential or exponential kernel.

Finally, the delta method is a classic forecasting approach that typically improves over standard averaging. For this method, we calculate the change between the future year and the historical period for all locations $b$ for each climate model. We then average the deltas for each location (b) across the entire climate model ensemble to compute a model ``delta'' for each model. Finally, we shift the historical target fields by delta and take the median of the shifted fields as our prediction.

\subsection{Climate field prediction} \label{sec:point_estimation}
We quantify predictive accuracy with the mean squared error (MSE) and the Structural Similarity Index Measure (SSIM) \citep{wang2002universal}. MSE quantifies overall accuracy, and SSIM quantifies the visual similarity of the prediction with the target. SSIM overcomes a shortcoming of MSE: a low MSE is achievable even with predictions that do not resemble the target. High SSIM values indicate that structural features of the climate model, such as high precipitation events and orographic temperature gradients (Figure \ref{fig:prediction_viz}), are well preserved. Moreover, high SSIM means the predictions are not blurred, noisy, skewed, or otherwise distorted, even at very fine scales. We use the standard equal area-weighted MSE
\begin{equation*}
	MSE(y, \hat{y}) = \frac{1}{n}\sum_p w_p (y - \hat{y})^2,
\end{equation*}
where $w_p$ are latitude-based weights \citep{north1982sampling}, $y$ is the observation and $\hat{y}$ is the prediction. For SSIM, we use the standard, though complex, sliding window SSIM described in \citet{wang2002universal}.

\begin{table}[]
\centering
\resizebox{\textwidth}{!}{%
\begin{tabular}{c|cccccccc}
\multicolumn{8}{l}{($\downarrow$) Mean Squared Error (MSE) - T2M} \\
\toprule
Model & 2030 & 2040 & 2050 & 2060 & 2070 & 2080 & 2090 & 2100 \\
\midrule
NN-GPR & 1.91 (0.06) & 1.97 (0.06) & 2.10 (0.07) & 2.27 (0.08) & 2.37 (0.09) & 2.53 (0.11) & 2.68 (0.11) & 2.84 (0.12) \\
LM     & 2.29 (0.11) & 2.28 (0.10) & 2.38 (0.12) & 2.51 (0.13) & 2.54 (0.14) & 2.57 (0.17) & 2.62 (0.17) & 2.71 (0.19) \\
WEA    & 3.29 (0.22) & 3.27 (0.20) & 3.40 (0.23) & 3.54 (0.25) & 3.54 (0.25) & 3.60 (0.27) & 3.62 (0.28) & 3.67 (0.28) \\
EA     & 5.98 (0.53) & 5.87 (0.50) & 5.96 (0.49) & 6.04 (0.45) & 6.00 (0.45) & 6.03 (0.43) & 5.97 (0.43) & 5.99 (0.42) \\
GPSE   & 1.91 (0.06) & 2.01 (0.06) & 2.26 (0.08) & 2.57 (0.09) & 2.85 (0.12) & 3.23 (0.13) & 3.60 (0.15) & 3.96 (0.17) \\
GPEX   & 1.89 (0.06) & 1.97 (0.06) & 2.19 (0.07) & 2.44 (0.08) & 2.65 (0.10) & 2.90 (0.11) & 3.16 (0.11) & 3.40 (0.13) \\
CNN    & 2.78 (0.15) & 2.75 (0.14) & 2.79 (0.17) & 2.95 (0.18) & 2.94 (0.18) & 2.97 (0.22) & 3.01 (0.23) & 3.08 (0.24) \\
DELT   & 3.07 (0.22) & 3.05 (0.21) & 3.17 (0.23) & 3.31 (0.24) & 3.30 (0.23) & 3.36 (0.25) & 3.40 (0.25) & 3.46 (0.26) \\
\bottomrule
\multicolumn{1}{c}{} & \\
\multicolumn{8}{l}{($\uparrow$) Structural Similarity Index (SSIM) - T2M} \\
\toprule
Model & 2030 & 2040 & 2050 & 2060 & 2070 & 2080 & 2090 & 2100 \\
\midrule
NN-GPR & 0.92 (0.00) & 0.92 (0.00) & 0.91 (0.00) & 0.91 (0.00) & 0.90 (0.00) & 0.89 (0.00) & 0.89 (0.01) & 0.88 (0.01) \\
LM     & 0.91 (0.00) & 0.91 (0.00) & 0.91 (0.00) & 0.90 (0.00) & 0.90 (0.01) & 0.90 (0.01) & 0.89 (0.01) & 0.89 (0.01) \\
WEA    & 0.89 (0.01) & 0.88 (0.01) & 0.88 (0.01) & 0.88 (0.01) & 0.87 (0.01) & 0.87 (0.01) & 0.87 (0.01) & 0.87 (0.01) \\
EA     & 0.84 (0.01) & 0.84 (0.01) & 0.83 (0.01) & 0.83 (0.01) & 0.83 (0.01) & 0.82 (0.01) & 0.82 (0.01) & 0.82 (0.01) \\
GPSE   & 0.92 (0.00) & 0.91 (0.00) & 0.90 (0.00) & 0.89 (0.00) & 0.88 (0.01) & 0.87 (0.01) & 0.86 (0.01) & 0.86 (0.01) \\
GPEX   & 0.92 (0.00) & 0.92 (0.00) & 0.91 (0.00) & 0.90 (0.00) & 0.89 (0.00) & 0.88 (0.00) & 0.88 (0.01) & 0.87 (0.01) \\
CNN    & 0.89 (0.01) & 0.89 (0.01) & 0.89 (0.01) & 0.88 (0.01) & 0.88 (0.01) & 0.88 (0.01) & 0.88 (0.01) & 0.88 (0.01) \\
DELT   & 0.89 (0.00) & 0.89 (0.00) & 0.88 (0.00) & 0.88 (0.00) & 0.88 (0.01) & 0.87 (0.01) & 0.87 (0.01) & 0.87 (0.01) \\
\bottomrule
\multicolumn{1}{c}{} & \\
\multicolumn{8}{l}{($\downarrow$) Mean Squared Error (MSE) - PR} \\
\toprule
Model & 2030 & 2040 & 2050 & 2060 & 2070 & 2080 & 2090 & 2100 \\
\midrule
NN-GPR & 3.84 (0.24) & 3.97 (0.27) & 4.05 (0.25) & 4.28 (0.25) & 4.47 (0.29) & 4.59 (0.28) & 4.68 (0.28) & 4.83 (0.31) \\
LM     & 4.41 (0.26) & 4.58 (0.28) & 4.64 (0.27) & 4.84 (0.26) & 4.99 (0.29) & 5.08 (0.28) & 5.16 (0.30) & 5.29 (0.32) \\
WEA    & 4.97 (0.33) & 5.14 (0.35) & 5.22 (0.34) & 5.43 (0.34) & 5.58 (0.38) & 5.65 (0.37) & 5.76 (0.39) & 5.85 (0.39) \\
EA     & 5.84 (0.37) & 6.03 (0.39) & 6.13 (0.37) & 6.33 (0.37) & 6.48 (0.42) & 6.57 (0.41) & 6.70 (0.42) & 6.76 (0.43) \\
GPSE   & 3.88 (0.23) & 4.02 (0.26) & 4.08 (0.25) & 4.33 (0.24) & 4.52 (0.28) & 4.63 (0.27) & 4.73 (0.28) & 4.87 (0.30) \\
GPEX   & 3.86 (0.23) & 3.99 (0.26) & 4.06 (0.25) & 4.31 (0.25) & 4.49 (0.29) & 4.61 (0.27) & 4.71 (0.28) & 4.86 (0.31) \\
CNN    & 4.70 (0.28) & 4.87 (0.30) & 4.92 (0.29) & 5.15 (0.28) & 5.34 (0.32) & 5.41 (0.32) & 5.49 (0.33) & 5.63 (0.35) \\
DELT   & 5.15 (0.30) & 5.31 (0.33) & 5.40 (0.30) & 5.60 (0.31) & 5.74 (0.35) & 5.85 (0.33) & 5.97 (0.35) & 6.05 (0.36) \\
\bottomrule
\multicolumn{1}{c}{} & \\
\multicolumn{8}{l}{($\uparrow$) Structural Similarity Index (SSIM) - PR} \\
\toprule
Model & 2030 & 2040 & 2050 & 2060 & 2070 & 2080 & 2090 & 2100 \\
\midrule
NN-GPR & 0.59 (0.02) & 0.58 (0.02) & 0.58 (0.02) & 0.57 (0.02) & 0.56 (0.02) & 0.56 (0.02) & 0.56 (0.02) & 0.55 (0.02) \\
LM     & 0.55 (0.02) & 0.55 (0.02) & 0.54 (0.02) & 0.54 (0.02) & 0.54 (0.02) & 0.54 (0.02) & 0.53 (0.02) & 0.53 (0.02) \\
WEA    & 0.50 (0.02) & 0.49 (0.02) & 0.49 (0.02) & 0.49 (0.02) & 0.49 (0.02) & 0.49 (0.02) & 0.48 (0.02) & 0.48 (0.02) \\
EA     & 0.48 (0.02) & 0.47 (0.02) & 0.47 (0.02) & 0.47 (0.02) & 0.47 (0.02) & 0.47 (0.02) & 0.46 (0.02) & 0.47 (0.02) \\
GPSE   & 0.58 (0.02) & 0.58 (0.02) & 0.57 (0.02) & 0.56 (0.02) & 0.55 (0.02) & 0.55 (0.02) & 0.54 (0.02) & 0.54 (0.02) \\
GPEX   & 0.58 (0.02) & 0.58 (0.02) & 0.57 (0.02) & 0.56 (0.02) & 0.55 (0.02) & 0.55 (0.02) & 0.54 (0.02) & 0.54 (0.02) \\
CNN    & 0.51 (0.02) & 0.51 (0.02) & 0.51 (0.02) & 0.50 (0.02) & 0.50 (0.02) & 0.50 (0.02) & 0.49 (0.02) & 0.49 (0.02) \\
DELT   & 0.51 (0.02) & 0.51 (0.02) & 0.51 (0.02) & 0.51 (0.02) & 0.50 (0.02) & 0.50 (0.02) & 0.50 (0.02) & 0.50 (0.02) \\
\bottomrule
\end{tabular}}
\caption{Average MSE and SSIM for eight model integration approaches broken out by the eight decades in the testing period. Top two tables show \textbf{surface temperature} (T2M) results, while the bottom two tables show \textbf{total precipitation} (PR) results. Arrows indicate whether higher ($\uparrow$) or lower ($\downarrow$) numbers are better. Decade numbers indicate the past decade up to that year, e.x. 2030 means error calculated from 2021-2030 data, and 2040 means from 2031-2040, etc.}
\label{tab:point_metrics}
\end{table}

\begin{figure}
	\centering
	\includegraphics[width=0.9\textwidth]{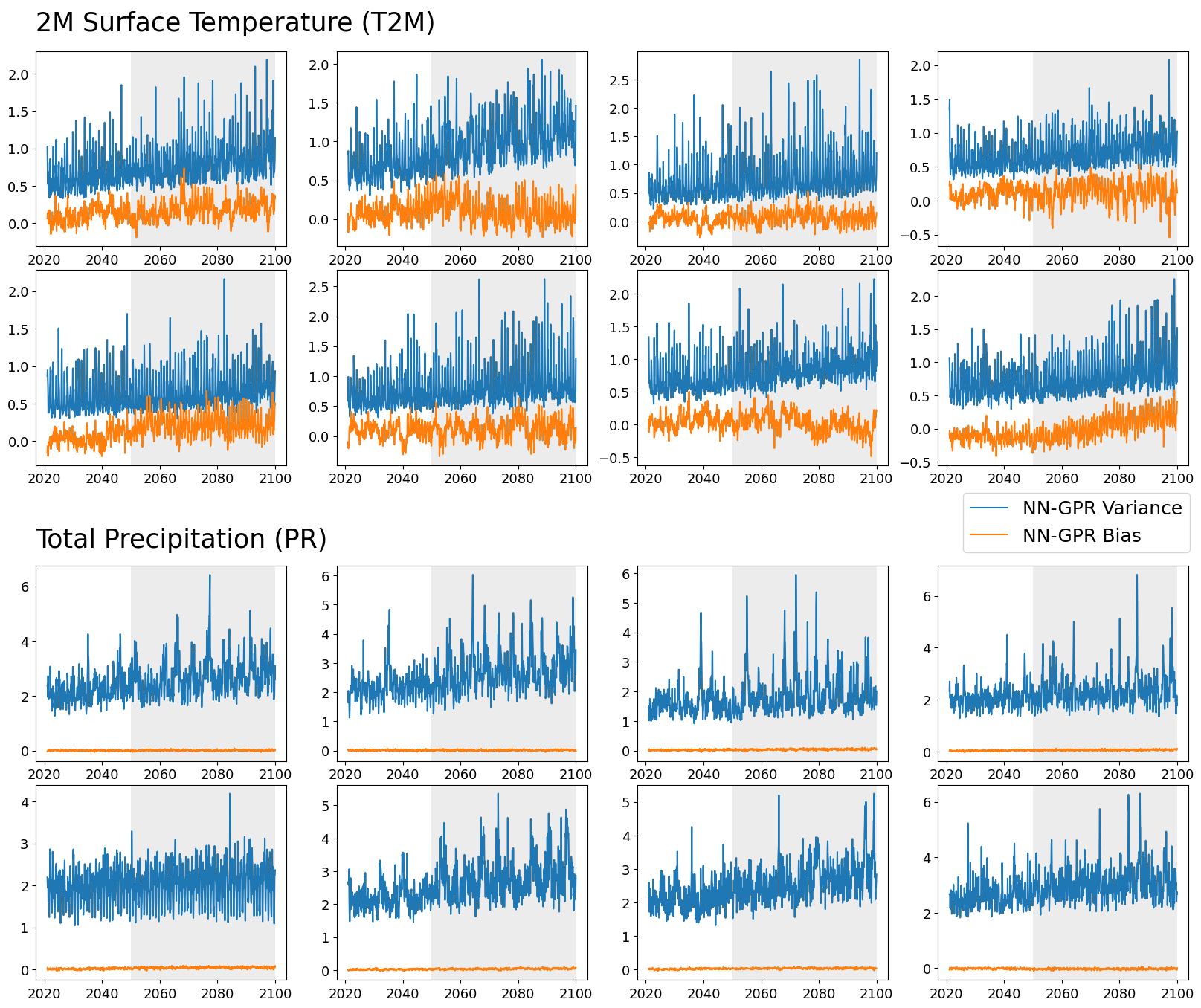}
    \caption{Bias-variance decomposition of NN-GPR's prediction MSE for 8 of the 16 models in the ensemble for T2M (top) and PR (bottom). All predictions are made on the test set using SSP2-4.5 forced model simulations, as in Table \ref{tab:point_metrics}. NN-GPR tends to show low bias and increasing variance over the prediction interval.}
    \label{fig:bias_variance}
\end{figure}

Table \ref{tab:point_metrics} shows the average (over the 16 perfect model tests) of the MSE and SSIM scores for each method applied to temperature (T2M) and precipitation (PR) prediction. For T2M, NN-GPR strongly improves over EA, WEA, CNN, and DELT over the entire prediction interval in terms of MSE and SSIM. NN-GPR also improves over GPSE and GPEX, though the differences are slight between 2030-2060. Against LM, NN-GPR is a strong improvement in the first five decades but deteriorates to a similar performance in the last three decades (2070-2100). The reason is because of the way NN-GPR manages bias and variance (Figure \ref{fig:bias_variance}). For PR, NN-GPR is a clear improvement over LM, WEA, EA, CNN, and DELT and only negligibly improves over GPSE and GPEX. We explore the differences between NN-GPR and GPSE more closely in the Appendix (Section B.3.) and find that while their summary metrics are close, NN-GPR consistently shows lower MSE and higher SSIM scores across time and model experiment.

Figure \ref{fig:bias_variance} investigates why NN-GPR's performance deteriorates in the latter period (2070-2100) for T2M and not for PR. NN-GPR shows low bias over the entire prediction interval, but the variance trends upwards. Over longer time horizons (2050-2100 shaded in grey), the variance steadily increases, leading to inflated MSE values. However, the bias remains stable around zero. Thus, while the NN-GPR may produce marginally suboptimal predictions in the long term (2080 - 2100), we expect the spatial averages of those predictions, such as global mean surface temperatures, to be relatively unbiased.

\begin{figure}
	\centering
  	\includegraphics[width=0.8\textwidth]{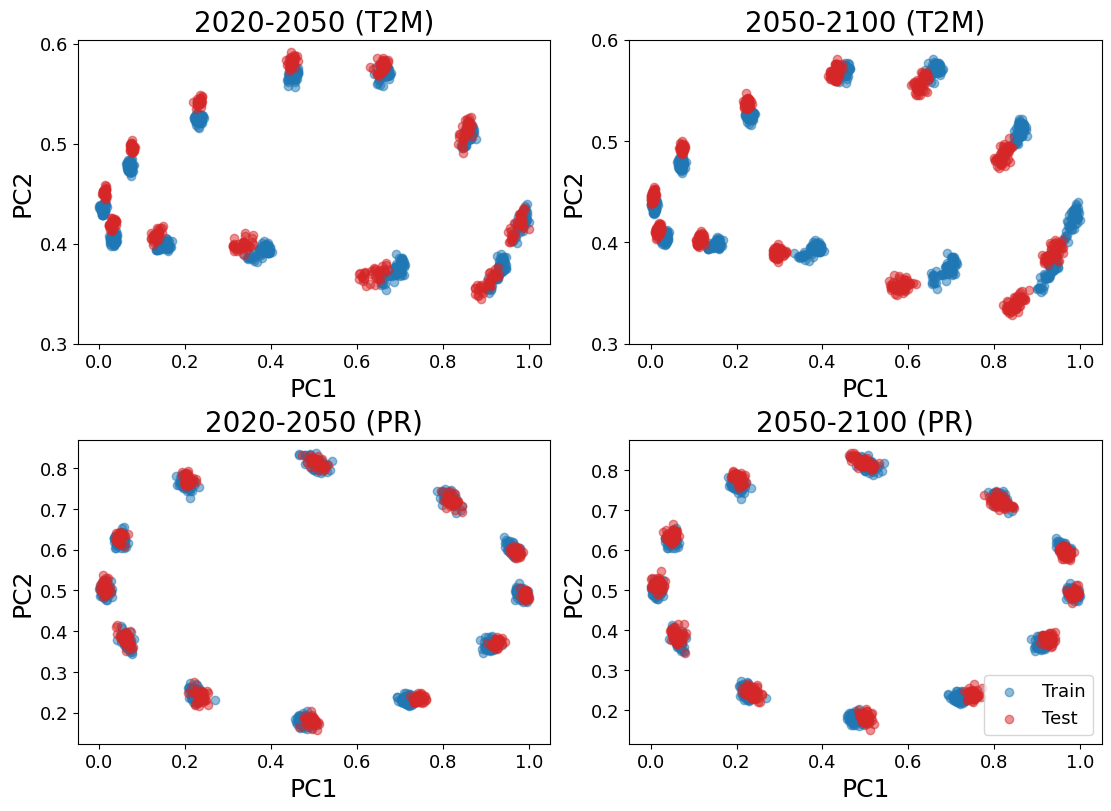}
  	\caption{Projections of the historical climate model data and the future SSP model simulations onto the top two principal components of the historical model data. Model means and variances were removed before projection. The top row shows projections of T2M data, and the bottom row shows PR data, divided into the near-term (2020-2050) and long-term (2050-2100) sections of the prediction period.}
\label{fig:covariate_shift}
\end{figure}

The variance trends upwards because the SSP2-4.5 models (inputs) become more and more unlike the historical climate models. That is, our T2M predictions suffer from \textit{covariate shift} \citep{shimodaira2000improving} (Figure \ref{fig:covariate_shift}a). NN-GPR predictions based on inputs (climate models) that are increasingly dissimilar from the training data converge to the same weighted average of the training reanalysis fields plus the trend. After further analysis, we found that this resulted in over smoothing of the distinctive features of the predicted reanalysis fields. Thus, the prediction error variance and SSIM increased, but the bias remained low. T2M shows strong evidence of covariate shift due to the mismatch between the training and test point clouds and, thus, higher variance inflation over time. 

However, PR does not experience the same degree of performance deterioration as T2M does since PR does not experience the same covariate shift as T2M (Figure \ref{fig:covariate_shift}b). We expect a substantial covariate shift in T2M due to climate change, but the shift for precipitation is small relative to the monthly averages. Thus, total monthly precipitation fields under SSP2-4.5 in the far future are similar to the training data, so variance and SSIM do not necessarily inflate and deflate, respectively.

These results indicate that NN-GPR can approximate individual model simulations and improve over highly parameterized models such as pointwise linear regression. Thus NN-GPR can be a highly effective approach to climate model integration. However, its improved prediction skill is tied to our ability to mitigate covariate shift, such as by including an explicit trend term. In light of Figures \ref{fig:bias_variance} and \ref{fig:covariate_shift}, we can partially explain the discrepancy between NN-GPR and GPSE / GPEX in Table \ref{tab:point_metrics} for T2M, but not for PR, as NN-GPR being, evidently, more robust to covariate shift. Figure \ref{fig:covariate_shift} demonstrates that covariate shift is not an inherent feature of time but the variable type. Thus, any climate model integration needs to consider, on a case-by-case basis, how to quantify and account for covariate shift, i.e., simulated climate change present in the future climate model ensemble.

\subsection{Uncertainty quantification} \label{sec:uncertainty_quantification}
We compare each method's uncertainty quantification (UQ) abilities through two metrics. The first is the Continuous Ranked Probability Score (CRPS), a proper scoring rule used commonly used to assess probabilistic predictions in climate science \citep{gneiting2007strictly}. The second is the Euclidean distance between the probability integral transform of each method's prediction and a uniform distribution (PIT). CRPS measures a combination of calibration and sharpness, while PIT explicitly focuses on calibration \citep{gneiting2007probabilistic}. As an empirical Bayesian method, NN-GPR produces credible intervals rather than confidence intervals, so calibration is not guaranteed. We exclude CNN from UQ comparisons since it does not have a natural UQ mechanism.

\begin{table}[]
\centering
\resizebox{\textwidth}{!}{%
\begin{tabular}{c|cccccccc}
\multicolumn{8}{l}{($\downarrow$) Continuous Ranked Probability Score (CRPS) - T2M} \\
\toprule
Model & 2030 & 2040 & 2050 & 2060 & 2070 & 2080 & 2090 & 2100 \\
\midrule
NN-GPR & 0.73 (0.01) & 0.74 (0.01) & 0.76 (0.01) & 0.79 (0.01) & 0.81 (0.01) & 0.83 (0.02) & 0.86 (0.02) & 0.88 (0.02) \\
LM     & 0.68 (0.02) & 0.69 (0.02) & 0.69 (0.02) & 0.72 (0.02) & 0.73 (0.02) & 0.74 (0.02) & 0.74 (0.02) & 0.76 (0.02) \\
WEA    & 1.15 (0.05) & 1.15 (0.05) & 1.16 (0.05) & 1.18 (0.05) & 1.17 (0.04) & 1.18 (0.04) & 1.18 (0.04) & 1.18 (0.04) \\
EA     & 1.15 (0.05) & 1.15 (0.05) & 1.16 (0.05) & 1.18 (0.05) & 1.17 (0.04) & 1.18 (0.04) & 1.18 (0.04) & 1.18 (0.04) \\
GPSE   & 0.73 (0.01) & 0.74 (0.01) & 0.77 (0.01) & 0.81 (0.01) & 0.84 (0.01) & 0.88 (0.02) & 0.92 (0.02) & 0.94 (0.02) \\
GPEX   & 0.73 (0.01) & 0.75 (0.01) & 0.78 (0.01) & 0.82 (0.01) & 0.86 (0.02) & 0.92 (0.02) & 0.97 (0.02) & 1.02 (0.02) \\
DELT   & 3.87 (0.04) & 3.93 (0.04) & 4.00 (0.04) & 4.06 (0.04) & 4.12 (0.04) & 4.16 (0.04) & 4.21 (0.04) & 4.24 (0.04) \\
\bottomrule
\multicolumn{1}{c}{} & \\
\multicolumn{8}{l}{($\downarrow$) Probability Integral Transform (PIT) - T2M} \\
\toprule
Model & 2030 & 2040 & 2050 & 2060 & 2070 & 2080 & 2090 & 2100 \\
\midrule
NN-GPR & 0.59 (0.02) & 0.57 (0.01) & 0.58 (0.01) & 0.56 (0.02) & 0.55 (0.02) & 0.53 (0.02) & 0.53 (0.02) & 0.52 (0.02) \\
LM     & 0.26 (0.01) & 0.28 (0.02) & 0.31 (0.02) & 0.36 (0.03) & 0.37 (0.03) & 0.41 (0.04) & 0.42 (0.04) & 0.44 (0.04) \\
WEA    & 0.40 (0.02) & 0.40 (0.02) & 0.42 (0.03) & 0.44 (0.03) & 0.45 (0.03) & 0.45 (0.04) & 0.45 (0.04) & 0.45 (0.04) \\
EA     & 0.46 (0.05) & 0.47 (0.04) & 0.48 (0.05) & 0.50 (0.04) & 0.51 (0.05) & 0.51 (0.05) & 0.50 (0.05) & 0.50 (0.05) \\
GPSE   & 0.58 (0.02) & 0.56 (0.02) & 0.55 (0.01) & 0.52 (0.01) & 0.49 (0.02) & 0.46 (0.02) & 0.44 (0.02) & 0.43 (0.02) \\
GPEX   & 0.56 (0.02) & 0.53 (0.01) & 0.52 (0.01) & 0.48 (0.01) & 0.46 (0.02) & 0.45 (0.02) & 0.45 (0.02) & 0.45 (0.02) \\
DELT   & 1.58 (0.01) & 1.58 (0.01) & 1.58 (0.02) & 1.58 (0.02) & 1.59 (0.02) & 1.60 (0.02) & 1.61 (0.02) & 1.61 (0.02) \\
\bottomrule
\multicolumn{1}{c}{} & \\
\multicolumn{8}{l}{($\downarrow$) Continuous Ranked Probability Score - PR} \\
\toprule
Model & 2030 & 2040 & 2050 & 2060 & 2070 & 2080 & 2090 & 2100 \\
\midrule
NN-GPR & 0.92 (0.02) & 0.93 (0.03) & 0.94 (0.02) & 0.95 (0.02) & 0.98 (0.03) & 0.99 (0.02) & 1.00 (0.02) & 1.01 (0.03) \\
LM     & 0.89 (0.02) & 0.91 (0.02) & 0.91 (0.02) & 0.92 (0.02) & 0.94 (0.02) & 0.94 (0.02) & 0.95 (0.02) & 0.96 (0.02) \\
WEA    & 1.00 (0.03) & 1.02 (0.03) & 1.02 (0.03) & 1.03 (0.03) & 1.04 (0.03) & 1.05 (0.03) & 1.06 (0.03) & 1.06 (0.03) \\
EA     & 1.00 (0.03) & 1.02 (0.03) & 1.02 (0.03) & 1.03 (0.03) & 1.04 (0.03) & 1.05 (0.03) & 1.06 (0.03) & 1.06 (0.03) \\
GPSE   & 0.91 (0.02) & 0.93 (0.02) & 0.93 (0.02) & 0.94 (0.02) & 0.97 (0.02) & 0.98 (0.02) & 0.99 (0.02) & 1.00 (0.02) \\
GPEX   & 0.92 (0.02) & 0.93 (0.02) & 0.94 (0.02) & 0.95 (0.02) & 0.97 (0.02) & 0.98 (0.02) & 1.00 (0.02) & 1.00 (0.02) \\
DELT   & 0.66 (0.01) & 0.66 (0.01) & 0.67 (0.01) & 0.67 (0.01) & 0.68 (0.01) & 0.68 (0.01) & 0.68 (0.01) & 0.68 (0.01) \\
\bottomrule
\multicolumn{1}{c}{} & \\
\multicolumn{8}{l}{($\downarrow$) Probability Integral Transform (PIT) - PR} \\
\toprule
Model & 2030 & 2040 & 2050 & 2060 & 2070 & 2080 & 2090 & 2100 \\
\midrule
NN-GPR & 0.78 (0.02) & 0.77 (0.02) & 0.76 (0.02) & 0.75 (0.02) & 0.74 (0.02) & 0.73 (0.02) & 0.72 (0.02) & 0.71 (0.02) \\
LM     & 0.26 (0.01) & 0.26 (0.01) & 0.26 (0.01) & 0.25 (0.01) & 0.25 (0.01) & 0.25 (0.01) & 0.25 (0.01) & 0.25 (0.01) \\
WEA    & 0.34 (0.01) & 0.33 (0.01) & 0.34 (0.01) & 0.34 (0.01) & 0.34 (0.01) & 0.34 (0.01) & 0.34 (0.01) & 0.34 (0.01) \\
EA     & 0.36 (0.01) & 0.36 (0.01) & 0.36 (0.01) & 0.36 (0.01) & 0.36 (0.01) & 0.36 (0.01) & 0.36 (0.01) & 0.36 (0.01) \\
GPSE   & 0.77 (0.02) & 0.76 (0.02) & 0.76 (0.02) & 0.75 (0.02) & 0.74 (0.02) & 0.73 (0.02) & 0.73 (0.02) & 0.73 (0.02) \\
GPEX   & 0.75 (0.02) & 0.75 (0.02) & 0.75 (0.02) & 0.74 (0.02) & 0.74 (0.02) & 0.73 (0.02) & 0.72 (0.02) & 0.72 (0.02) \\
DELT   & 0.59 (0.02) & 0.59 (0.02) & 0.59 (0.02) & 0.59 (0.02) & 0.59 (0.02) & 0.59 (0.02) & 0.59 (0.02) & 0.59 (0.02) \\
\bottomrule
\end{tabular}}
\caption{Average CRPS and PIT score for the eight different model integration approaches, broken out by the 8 decades in the testing period. Top two tables show \textbf{surface temperature} (T2M) results, while the bottom two tables show \textbf{total precipitation} (PR) results. CNN excluded due to lack of natural uncertainty quantification mechanism. Arrows indicate whether higher ($\uparrow$) or lower ($\downarrow$) numbers are better.}
\label{tab:interval_metrics}
\end{table}

Table \ref{tab:interval_metrics} shows that NN-GPR has comparable CRPS with LM, GPSE, and GPEX for T2M, with LM slightly lower than NN-GPR and GPSE and GPEX slightly higher. NN-GPR, GPSE, and GPEX are nearly identical in CPRS for PR, although WEA and DELT clearly dominate any of the Gaussian process methods here. The PIT comparisons reveal that, except for DELT on T2M, NN-GPR tends to be the least calibrated of all the approaches. This could be due to a variety of factors, such as NN-GPR not having a spatially varying variance term and the ReLU activation leading to overconfidence \citep{kristiadi2020being}. Without a spatially varying predictive variance, NN-GPR may be producing overly wide prediction intervals in some locations and overly narrow intervals in others.

\begin{figure}
	\centering
	\includegraphics[width=0.9\textwidth]{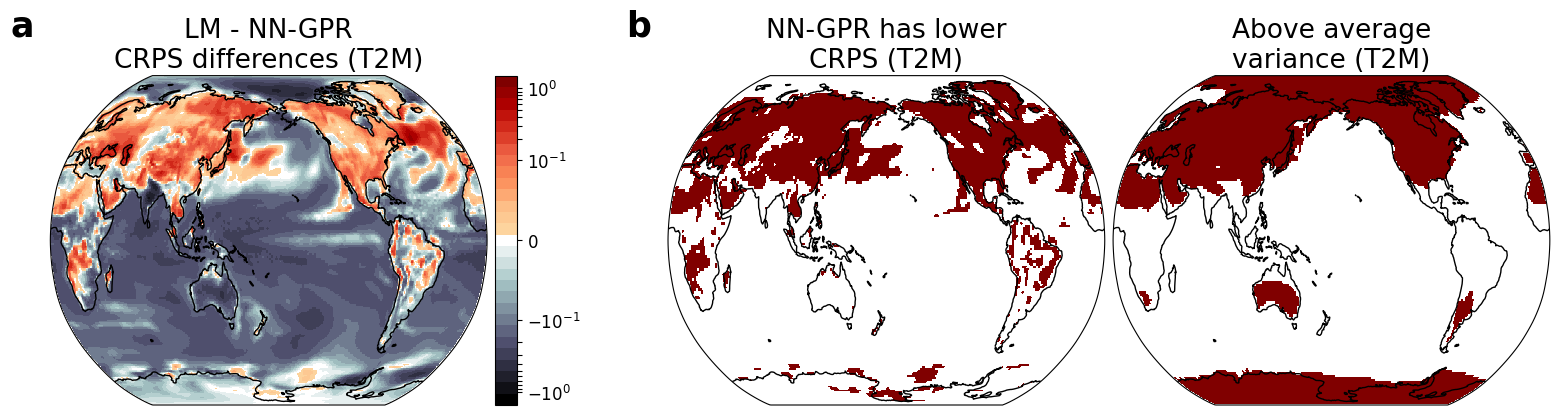} \\
	\includegraphics[width=0.9\textwidth]{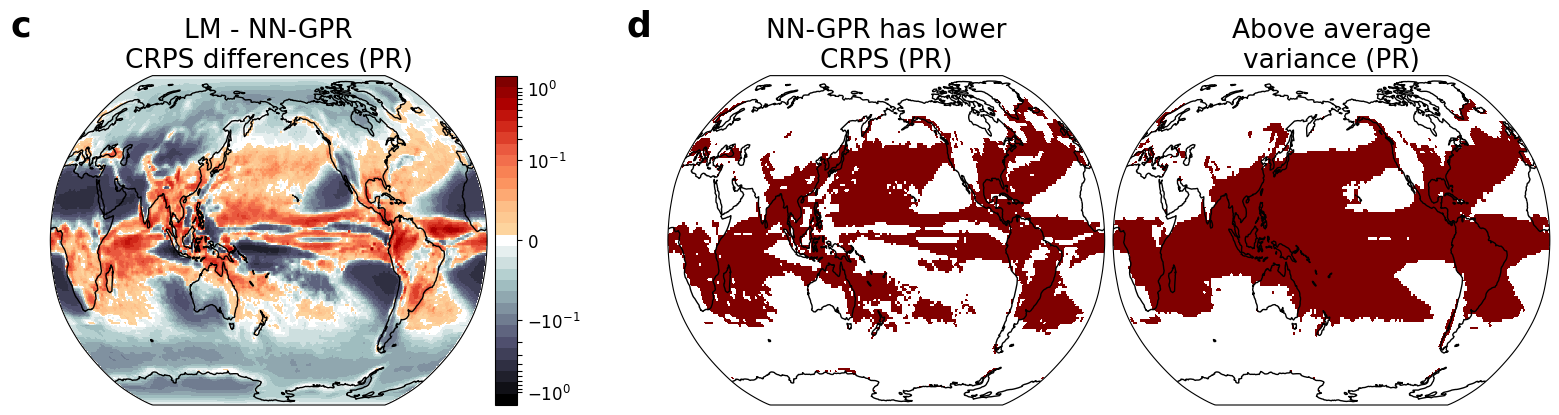} 
  	\caption{Panel (a) and (c) -- Average difference in CRPS scores (LM - NN-GPR) for a single example climate model (ACCESS-CM2). Orange regions indicate NN-GPR had lower CRPS than LM, while blue-grey indicates NN-GPR had higher CRPS than LM. Panel (b) and (d) -- The leftmost plots show a binarized version of panels (a) and (c) to highlight the exact regions where NN-GPR has lower CRPS scores than LM. The rightmost plots show regions in the ACCESS-CM2 output where the variance was higher than average}
    \label{fig:crps_diff}
\end{figure}

To see why and where miscalibration and incorrect sharpness occur, we created average CRPS maps for NN-GPR and LM by averaging their raw CRPS scores in time rather than in time and space as was done to create Table \ref{tab:interval_metrics}. We then subtracted NN-GPR's CRPS maps from LM's CRPS maps to produce Figure \ref{fig:crps_diff}a. This panel shows the map of the differences by averaging over the entire prediction period. For the T2M predictions, NN-GPR tends to have lower CRPS scores over land, particularly in the northern hemisphere, while LM shows lower CRPS scores over the ocean. A clear pattern emerges near the equator for the TP predictions where NN-GPR again significantly improves in CRPS over LM. To see why these patterns emerge in both variables, we created a second set of maps in (Figure \ref{fig:crps_diff}b, \ref{fig:crps_diff}d). These panels show locations where NN-GPR has lower CRPS than LM (improvements) and locations where the reanalysis fields exhibit higher than an average variance. Figures \ref{fig:crps_diff}b and \ref{fig:crps_diff}d indicate NN-GPR represents the climate model distribution better in regions of high variability. In areas of low variability, NN-GPR tends to overestimate the prediction intervals, leading to high CRPS and and PIT scores.

Combining the results from Table \ref{tab:point_metrics} with the maps in Figure \ref{fig:crps_diff}, we conclude that NN-GPR is generally more accurate than existing methods, but has relatively poorer UQ skill. Overly wide prediction intervals in regions of lower variability lead to the higher CRPS and higher PIT scores seen in Table \ref{tab:interval_metrics}. However, we also observed that NN-GPR represents variability in high variance regions better than the other methods (Figure \ref{fig:crps_diff}). This means optimizing the variance term in NN-GPR favors representing high-variance regions at the cost of being under-confident in low-variance regions since all regions are constrained to share a single variance value.

\section{Global and regional projections under SSP2-4.5} \label{sec:application}

Global climate models are essential tools for understanding how Earth's climate changes and for projecting regional impacts. Data products, such as the Coupled Model Intercomparison Project (CMIP6) \citep{eyring2016overview}, exist explicitly to study the projections from different climate models under a wide range of external forcings and socioeconomic scenarios. We again consider the two most widely studied climatic variables: temperature and precipitation. There is broad agreement between climate models that global temperatures are rising and that precipitation will become more intense and less frequent in the future \citep{trenberth2011changes, giorgi2019response}. However, the models can disagree significantly on how that process unfolds in time and space. 

We use NN-GPR to project future surface temperature and precipitation under Shared Socioeconomic Pathway 2 and RCP 4.5 (SSP2-4.5) \citep{o2016scenario}. We use the 16-member CMIP6 climate model ensemble (described in Section \ref{sec:data_models}) as input and predict reanalysis fields (described in Section \ref{sec:data_reanalysis}). We fit separate NN-GPR models for predicting 2-Meter Surface Temperature fields (T2M) and the Total Precipitation fields (PR) on single pressure levels. We quantitatively compare our NN-GPR against the ensemble average (EA) and pointwise linear model (LM) on their predictions for 2015-2021. Then, because there is no ground truth after 2021, we qualitatively compare our results against EA and LM thereafter. 

\subsection{Global climate projections}

\begin{figure}
	\centering
	\includegraphics[width=0.95\textwidth]{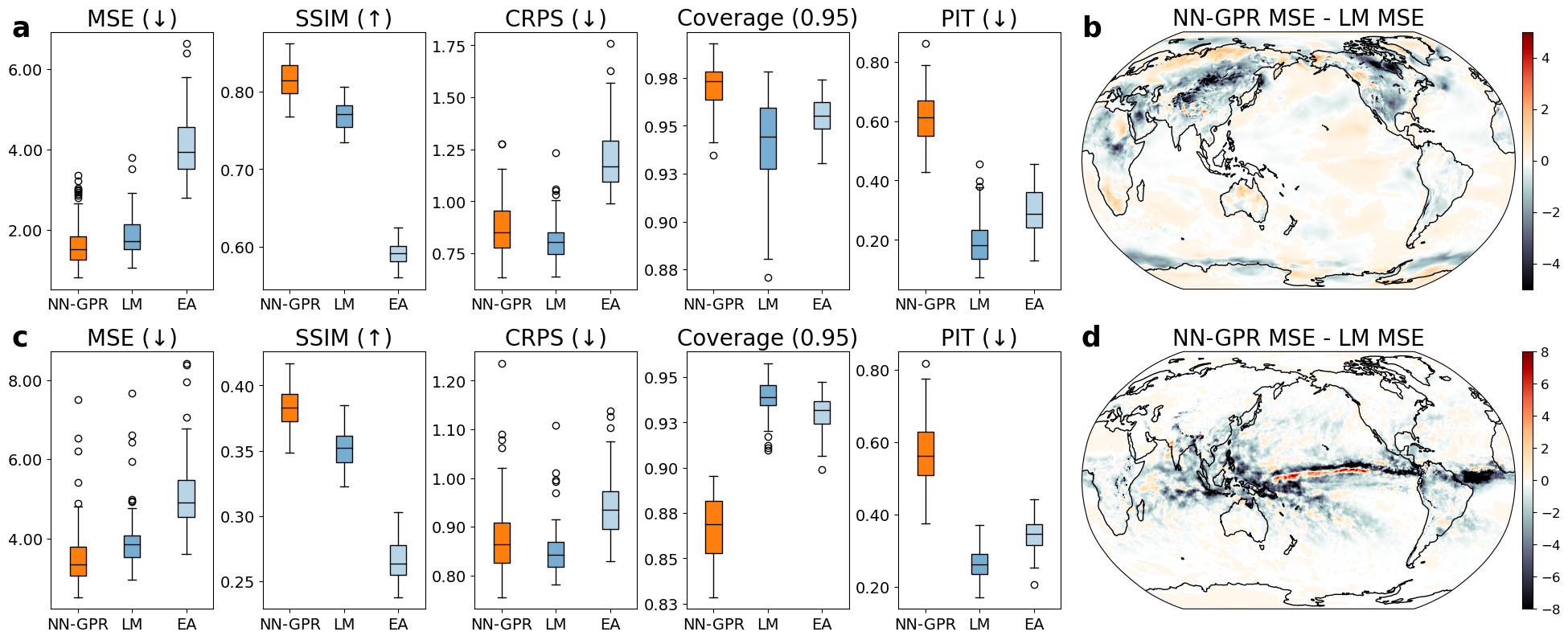} 
  	\caption{Boxplots comparing the MSE, SSIM, CRPS, 95\% Coverage, and PIT of each method over time for T2M (a) and PR (c). Boxplots treat each monthly prediction as an individual observation. Arrows indicate whether higher ($\uparrow$) or lower ($\downarrow$) numbers are better. Panels (b) and (d) show spatial differences in the MSE (averaged over time) of NN-GPR and LM for T2M(b) and PR(d).}
  	\label{fig:reanalysis_metrics}
\end{figure}

We first evaluate our method against LM and EA on reanalysis data to assess if there are any systematic differences between predicting models (Section \ref{sec:experiments}) and predicting reanalysis fields. We train each of the three methods using a subset of the historical data (1950-2015) and predict reanalysis fields from 2015-2021. We summarize the out-of-sample accuracy and uncertainty quantification metrics in Figure \ref{fig:reanalysis_metrics}.

Figures \ref{fig:reanalysis_metrics}a, \ref{fig:reanalysis_metrics}b shows the MSE, SSIM, CRPS, empirical coverage, and PIT of NN-GPR against LM and EA on temperature (T2M) and precipitation (PR) over the test period (2015-2021). The results are consistent with the experiments (Section \ref{sec:experiments}, Tables \ref{tab:point_metrics} and \ref{tab:interval_metrics}), with NN-GPR showing improved average prediction metrics (MSE: 1.69, SSIM: 0.82) compared to LM (MSE: 1.88, SSIM: 0.77) and EA (MSE: 4.07, SSIM: 0.59). Therefore, NN-GPR more accurately predicts the held-out historical observations and produces sharper, less distorted predictions. LM tends to have better uncertainty quantification metrics (CRPS: 0.82, Coverage: 0.94) than NN-GPR (CRPS: 0.88, Coverage: 0.97) since LM uses a spatially varying variance. NN-GPR again has relatively poor calibration (high PIT) compared to the other approaches.

\begin{figure}
	\centering
	\includegraphics[width=0.9\textwidth]{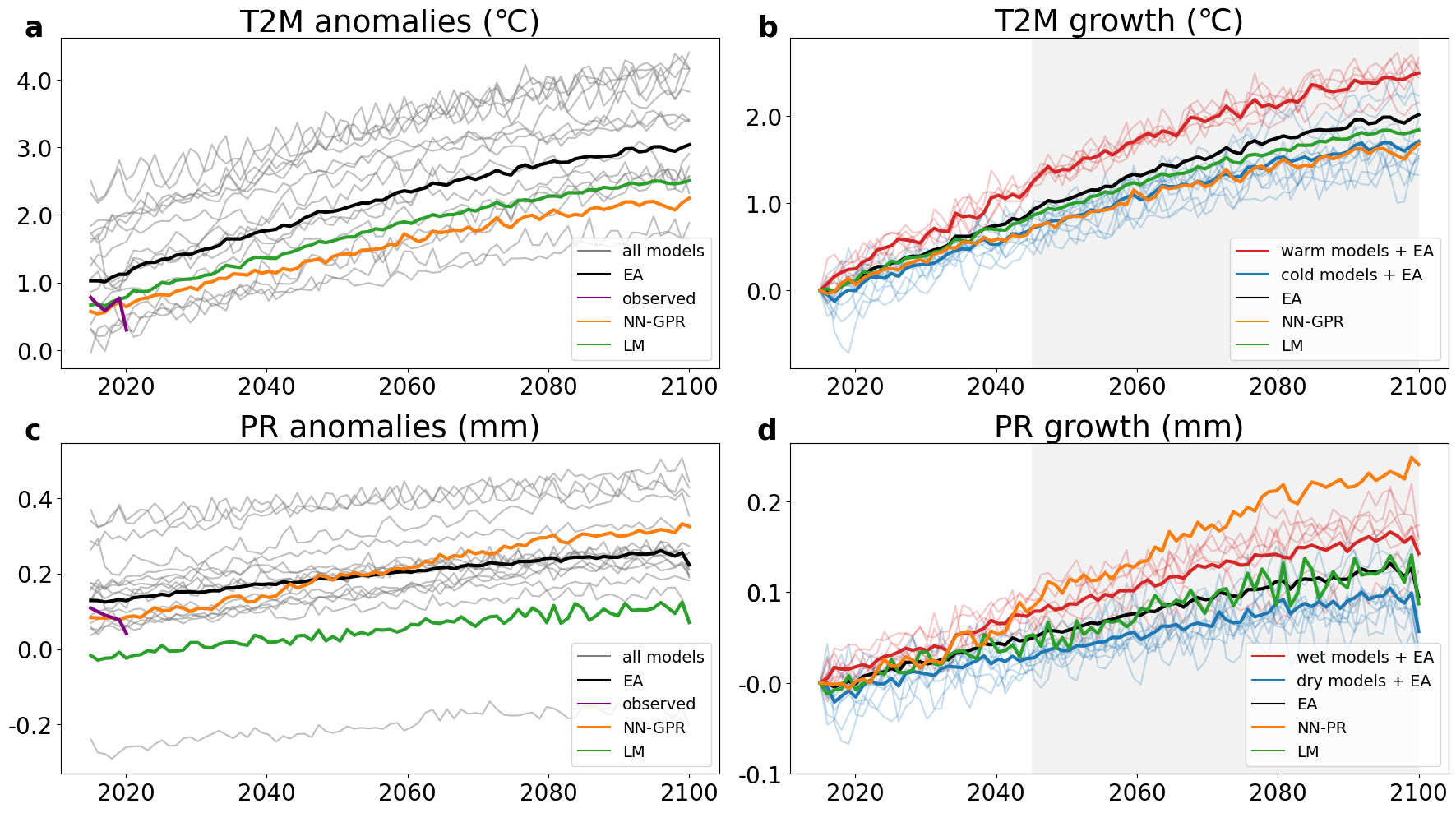} 
  	\caption{Panels (a) and (c) show yearly T2M and PR averages of each method, individual climate model averages (all models), and the observations. Panels (b) and (d) show projected T2M and PR growth of each method starting from 2015. In panel (b), a ``warm'' model has T2M growth over $1^\circ$C by 2045 and in panel (d) a ``wet'' model has PR growth over 0.05mm by 2045.}
  	\label{fig:reanalysis_projection}
\end{figure}

The MSE difference maps in Figures \ref{fig:reanalysis_metrics}b and \ref{fig:reanalysis_metrics}d illustrate where NN-GPR has higher historical skill than LM by showing regions where NN-GPR has a lower mean squared error. Overall, most locations have comparable MSE, although NN-GPR improves in regions of high variability as in Figure \ref{fig:crps_diff}. For T2M, these include high-latitude landmasses such as North America and Northern Asia. Improvements in PR largely stem from improvements in the Intertropical Convergence Zone (ITCZ) and southeast Asia (SEA). These findings are consistent with the numerical results in (Section \ref{sec:experiments}, Figure \ref{fig:crps_diff}).
Figure \ref{fig:reanalysis_metrics} supports our hypothesis that predictive skill in simulating climate model ensemble members jackknifed out of the ensemble (Section \ref{sec:experiments}) broadly translates to predictive skill on historical observation data, which can help reduce future projection uncertainty.

To see how each method's long-term projections (2021-2100) differ we predict from Jan. 2015 through Jan. 2099. We first compare the yearly global averages (T2M and PR) as estimated by each method (Figure \ref{fig:reanalysis_projection}). NN-GPR and LM both run cool compared to the model average ($0.67\, ^\circ$C  and $0.44\, ^\circ$C  below EA on average, respectively, in Figure \ref{fig:reanalysis_projection}a), though still within the model spread. NN-GPR produces similar global precipitation levels as EA (0.01mm higher than EA on average). At the same time, LM shows a considerably drier climate (0.15mm lower than EA on average) than all but a single ensemble member (Figure \ref{fig:reanalysis_projection}c). Figures \ref{fig:reanalysis_projection}a and \ref{fig:reanalysis_projection}c show that NN-GPR does not fundamentally alter the global averages beyond a slight shift and slope change to better fit the observed averages.

NN-GPR primarily differentiates itself from linear approaches in small scale, high variance regions (Figure \ref{fig:reanalysis_metrics}b, d), yet this has a profound impact on overall global patterns (Figures \ref{fig:reanalysis_projection}b and \ref{fig:reanalysis_projection}d). The T2M plot in Figure \ref{fig:reanalysis_projection}b shows that there are two groups of models in the data, with six of the models projecting rapid surface temperature increases around 2040, while the remaining ten steadily increase. NN-GPR closely tracks these cooler, steadier models and disregards the rapidly warming models, resulting in projected warming of $1.68\, ^\circ$C from 2020 to 2100 compared to $2.01\, ^\circ$C shown by the EA. Around 2045-2050, NN-GPR also begins projecting PR growth uncharacteristic of any ensemble model. This suggests that NN-GPR can produce long-range dynamics different from the climate model inputs.

\begin{figure}
	\centering
	\includegraphics[width=0.9\textwidth]{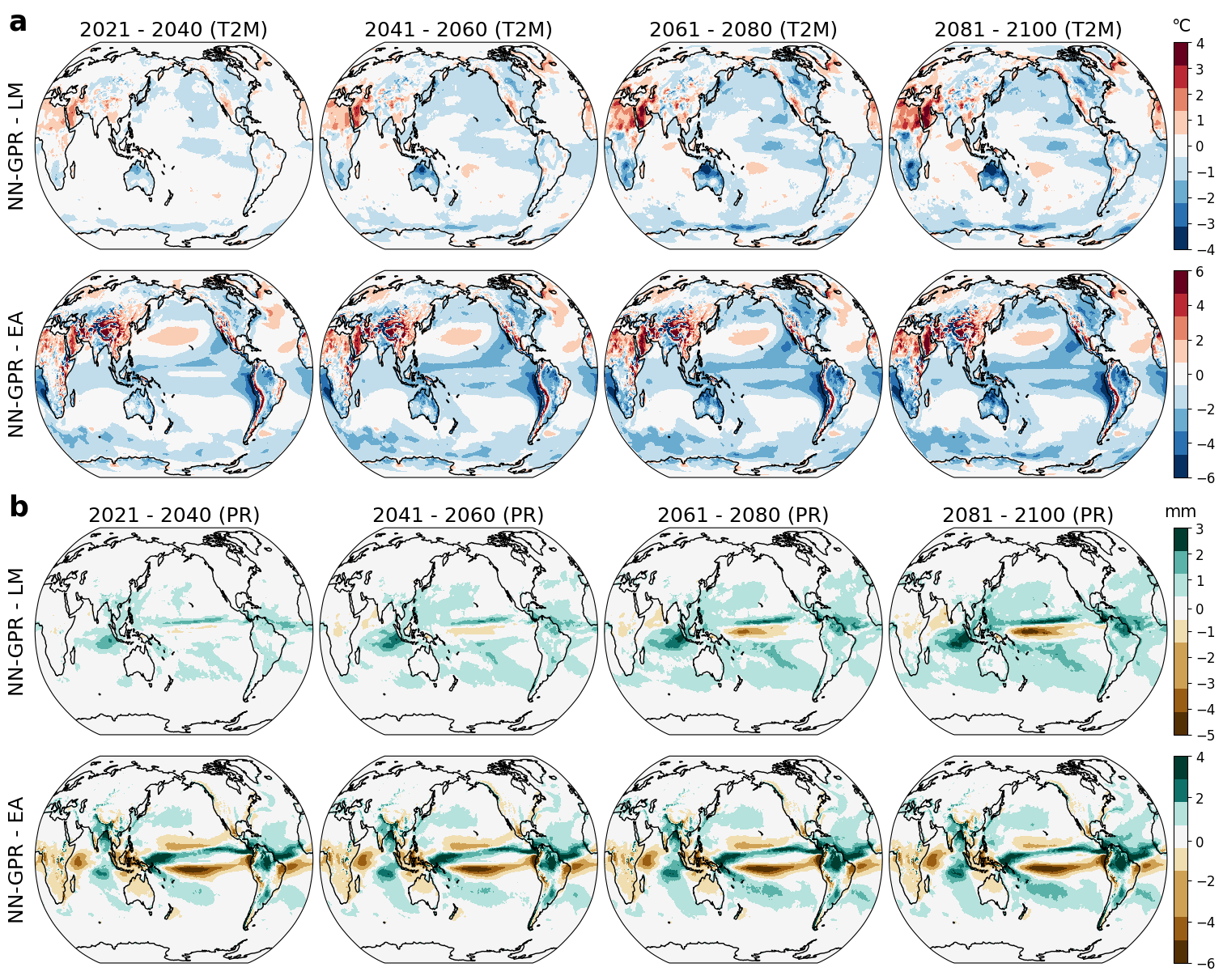} 
    \caption{Spatial differences between NN-GPR's projections and LMs and EA's projections averaged over two-decade blocks. Panel (a) -- T2M differences between NN-GPR and LM (top row) and NN-GPR and EA (bottom row). Red areas indicate NN-GPR projects higher temperatures, while blue areas indicate lower temperatures. Panel (b) -- PR differences between NN-GPR and LM (top row) and NN-GPR and EA (bottom row). Green areas mean NN-GPR projects higher precipitation, and brown areas mean NN-GPR projects lower precipitation.} 
  	\label{fig:reanalysis_spatial}
\end{figure}

Figure \ref{fig:reanalysis_spatial} shows the spatial differences between NN-GPR's projections and LMs and EA's projections to reveal why NN-GPR projects a relatively cooler and wetter future than either LM or EA. Panel \ref{fig:reanalysis_spatial}a shows that T2M differences are primarily driven by the Pacific Ocean, Southern Ocean, Australia, and North America, with Australia in particular much cooler (around $3\, ^\circ$C below the LM projection) during 2081-2100. NN-GPR - EA differences are broadly similar, with the west coast of South America additionally predicted to be much cooler than EA (around $4-6\, ^\circ$C below the EA projection). The Middle East / North Africa notably contradicts the general trend, with NN-GPR predicting around $2.5\, ^\circ$C hotter Sahara desert than LM during 2081-2100. NN-GPR also predicts Asia and the Andes mountains as relatively hotter.

Figure \ref{fig:reanalysis_spatial}b shows the relative differences in PR between NN-GPR's projections and LM and EA. The largest differences occur near the intertropical convergence zone (ITCZ) and the Indian Ocean, with large swaths of minor differences in the surrounding oceans compared to LM and EA. Almost no significant differences exist between any methods' projections outside the tropics. Within the tropics, NN-GPR projects significantly more precipitation than LM after 2060. Differences between NN-GPR and EA are relatively constant, except that NN-GPR gradually projects higher precipitation in northern South America and the South Pacific.

Notably, NN-GPR projects much lower average precipitation in the equatorial mid-Pacific than EA or LM. A drier mid-Pacific is consistent with the relatively increased precipitation seen elsewhere, particularly near southeast Asia and Northern South America, under La Ni\~{n}a like conditions \citep{lenssen2020seasonal}. Because we use a relatively small ensemble (16 members), some of the EA differences in PR may be due to the single outlying model (MCM-UA-1-0), as seen in Figure \ref{fig:reanalysis_projection}c.

\subsection{Regional climate projections} \label{sec:rcm_comparisons}

\begin{figure}
	\centering
	\includegraphics[width=0.9\textwidth]{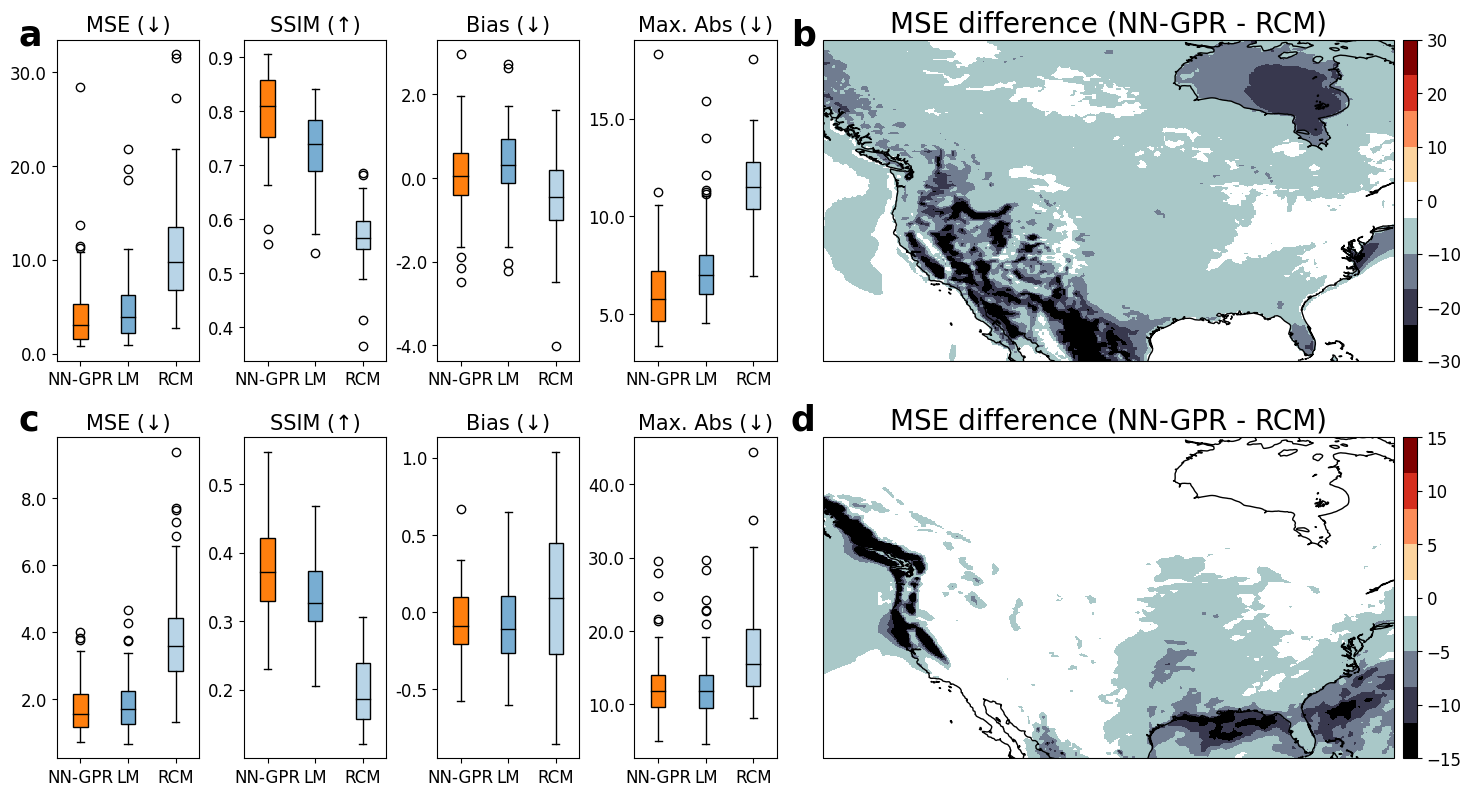} 
	\caption{Skill metrics for NN-GPR, LM and the regional climate model average (RCM) for T2M (a) and PR (c) forecasting over 2015-2021. Max. Abs refers to the absolute largest difference between the forecast and the reanalysis field. Arrows indicate whether higher ($\uparrow$) or lower ($\downarrow$) numbers are better. Panels (b) and (d) -- spatial MSE differences (NN-GPR - RCM) of T2M (a) and PR (c) projection.}
  	\label{fig:rcm_metrics}
\end{figure}

One of the critical benefits of NN-GPR is that low-resolution climate models are converted into a high-resolution prediction akin to statistical downscaling. To evaluate NN-GPR's downscaling effectiveness, we compare our regional predictions (based on global inputs) against LM predictions and the average of two regional climate models (RCMs), namely CanESM2 projections downscaled via CRCM5-OUR and CanRCM4.

Figures \ref{fig:rcm_metrics}a, c shows the MSE, the mean difference (Bias), the maximum absolute difference (Max. Abs), and SSIM scores of each method (NN-GPR, LM, and RCM) in predicting the test set (2015-2021) reanalysis fields for T2M and PR, respectively. NN-GPR has the lowest average MSE and highest average SSIM (MSE: 4.14, SSIM: 0.80) compared to the RCM (MSE: 10.92, SSIM: 0.57) and LM (MSE: 4.98, SSIM: 0.73). NN-GPR is, therefore, more accurate than either LM or RCM, meaning that it improves over both statistical and climatological downscaling approaches at a greatly reduced computational cost. The Bias and Max. Abs boxplots display the average and maximum absolute difference between the prediction and the observation. NN-GPR yields nearly unbiased estimates of the mean (Bias: $0.01\, ^\circ$C) and greatly improved extreme estimation (Max. Abs: $6.21\, ^\circ$C) compared to the RCM (Bias: $-0.43\, ^\circ$C, Max. Abs: $11.62\, ^\circ$C) and LM (Bias: $0.29\, ^\circ$C, Max. Abs: $7.51\, ^\circ$C). Thus, NN-GPR captures the target process's average and extremal behavior better than the other methods.

\begin{figure}
	\centering
	\includegraphics[width=0.9\textwidth]{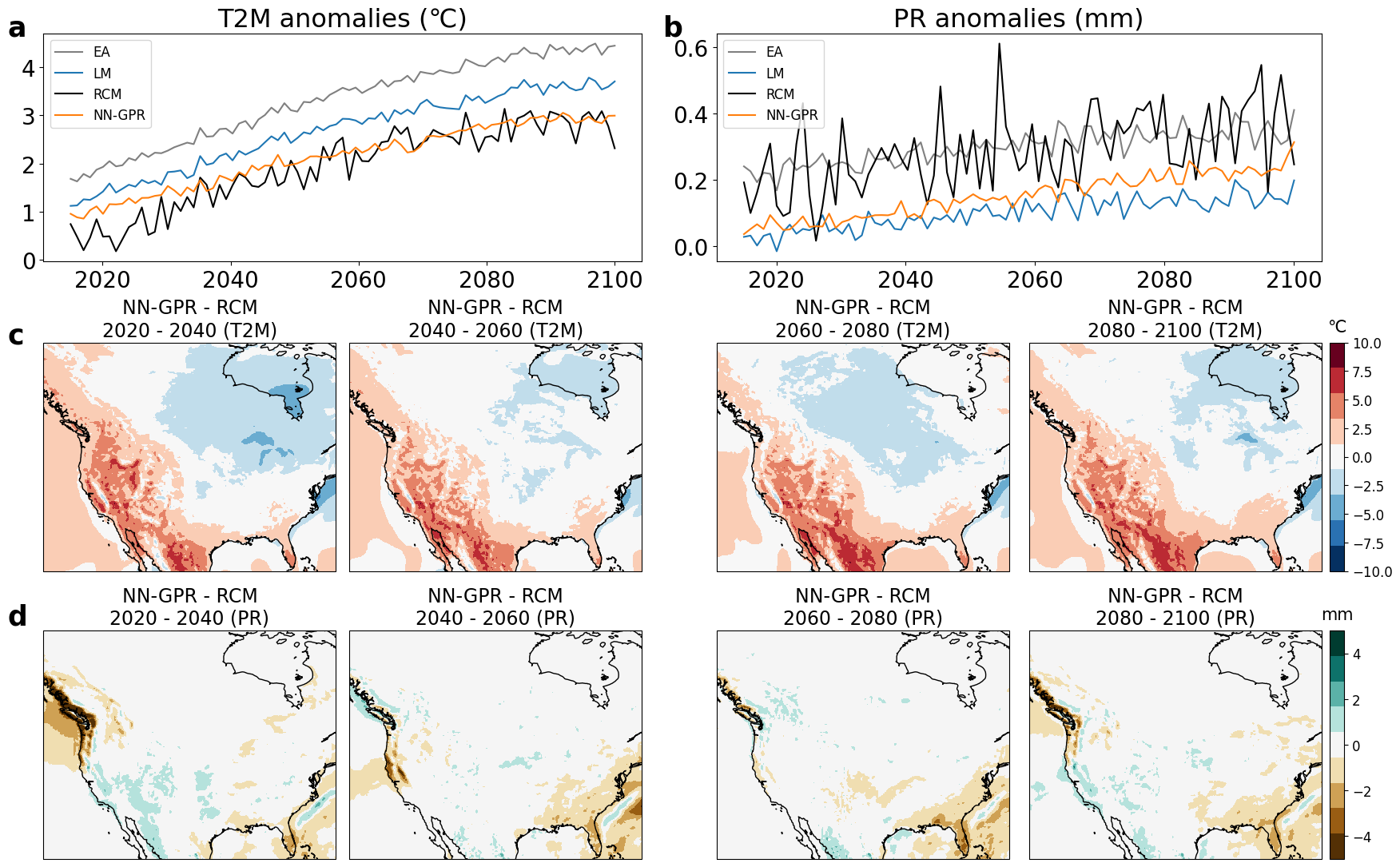} 
	\caption{Panels (a) and (b) -- yearly average anomaly projection (with respect to the 1950-2015 average) for each method and the RCM mean for T2M and PR. For T2M, NN-GPR closely tracks the RCM anomalies, while being consistently lower than RCM in PR. Panels (c) and (d) -- spatial differences between the NN-GPR and RCM mean for T2M (c) and PR (d).}
  	\label{fig:rcm_projections}
\end{figure}

Figures \ref{fig:rcm_projections}b, d show the temporally averaged MSE differences between NN-GPR and the RCM mean to show where NN-GPR provides improvement. NN-GPR has nearly uniformly smaller MSE values for T2M, with considerable improvements in the western United States and northern Mexico. Therefore, NN-GPR may capture interannual variability and topographic effects better than the RCM average. There are similarly large decreases in MSE in the Hudson Bay and the Chesapeake Bay for T2M in the Pacific Northwest and near Florida for PR. These widespread improvements suggest that NN-GPR can surpass the historical skill of RCMs, particularly in regions of high variability. Thus, NN-GPR may offer a practical and computationally inexpensive surrogate for RCM ensembles, which are often used to assess extreme behavior \citep{mearns2017cordex, haugen2018estimating}. Because there is significantly less tradeoff between spatial resolution and computational cost compared to RCMs, we could cheaply study extreme behavior at fine spatial resolutions, particularly if we used finer time scales than monthly time steps where extreme behavior is more meaningful.
 
Figures \ref{fig:rcm_projections}a, b show the yearly spatial averages of each method's predictions of North American T2M and PR fields. Strikingly, we see a convergence of the NN-GPR and RCM predictions around 2040, with near total agreement by 2070. Despite the considerable differences early on (Figure \ref{fig:rcm_metrics}), the two methods broadly agree near the end of the projection period. This sheds new light on the global predictions in Figure \ref{fig:reanalysis_projection} and suggests that NN-GPR may act as a global downscaling model in addition to a forecasting model. Comparing the NN-GPR predictions with the EA predictions in Figure \ref{fig:rcm_projections}a, we can again see that NN-GPR runs cool, consistent with Figure \ref{fig:reanalysis_projection}. For PR (Figure \ref{fig:rcm_projections}b), our model projects a slightly drier climate than either the RCM or EA, although the rate of precipitation increase is projected to be higher.

Figure \ref{fig:rcm_projections}c shows that the early differences between NN-GPR and the RCM might be due to NN-GPR projecting a warmer Pacific. After 2060, the differences between the two models stabilize (Figure \ref{fig:rcm_projections}a), albeit with NN-GPR consistently predicting a significantly warmer western United States and slightly cooler Canada and Hudson Bay. Differences in PR (Figure \ref{fig:rcm_projections}d) are essentially negligible save for in the Pacific Northwest and the Southeast, where NN-GPR projects a significantly drier climate. These differences largely account for the consistent gap between NN-GPR and the RCM in Figure \ref{fig:rcm_projections}b.

One of the critical benefits of our approach is that no RCM is required to produce predictions at RCM resolutions that are on par with RCM quality. RCMs introduce new uncertainties and biases, require subjective decision making, and have substantial computing requirements. Our NN-GPR method does not require any interpolation, does not average out the features of individual models, and is relatively cheap to compute, making it well suited to even larger ensembles than considered here. The results in Figure \ref{fig:rcm_metrics} show that NN-GPR is an adequate stand-in for RCM prediction and can even improve over small ensembles of RCMs for some climate variables like T2M. This could be beneficial for estimating climate impacts in regions with little RCM coverage.

\section{Discussion} \label{sec:discussion}

We proposed a Gaussian process regression (GPR) based on deep neural network kernels \citep{lee2017deep}, called NN-GPR, to improve spatiotemporal climate model integration. NN-GPR predicts reanalysis fields based on ensembles of climate model output, which constitutes an integration of the climate models. Because NN-GPR uses a deep neural network kernel, the predictions approximate the predictions of a fully trained deep neural network with an equivalent architecture. However, NN-GPR allows us to quantify predictive uncertainty with the GP posterior predictive distribution (Equation \ref{eqn:gpr_predictions}). Our NN-GPR approach radically departs from traditional model weighting integration schemes since we dispense with model weights and instead consider posterior predictive means as integrations. That is, we treat model integration as a prediction problem rather than a mixture weight estimation problem.

The simulation results in Sections \ref{sec:point_estimation} and \ref{sec:uncertainty_quantification} showed that our proposed NN-GPR approach was more accurate than ensemble averaging and regression. Table \ref{tab:point_metrics} showed that NN-GPR was most accurate early in the T2M prediction interval and slowly deteriorated after that. We found that this was due to covariate shift in the distribution of T2M model ensembles (Figure \ref{fig:covariate_shift}) under SSP2-4.5, causing the variance of the predictions to increase (Figure \ref{fig:bias_variance}). For PR, our method was consistently more accurate than the existing approaches. This was likely due to the significantly smaller covariate shift in the PR model ensembles. 

The reanalysis results in Section \ref{sec:application} showed that NN-GPR greatly improved over ensemble averaging and moderately improved over pointwise linear regression. NN-GPR was found to produce similar overall global and regional mean predictions as the ensemble mean but significantly improved the fine-scale detail, such as in mountainous regions (Figure \ref{fig:prediction_viz}, \ref{fig:rcm_metrics}). These numerous local improvements reduced MSE and increased SSIM globally, particularly in high-variance regions. Looking further at the local level, we found that NN-GPR projections were highly competitive with RCM projections. NN-GPR consistently estimated North American T2M and PR fields with higher accuracy and fidelity than the RCM average (Figure \ref{fig:rcm_metrics}).

Through extensive testing, we found that NN-GPR is sensitive to covariate shift (Figure \ref{fig:covariate_shift}). When the distribution of the inputs (climate model ensemble) in the prediction period is different from that in the training period, NN-GPR tends to be less accurate and have higher predictive variance. This can adversely impact long-range temperature projections, which exhibited moderate covariate shift in our experiments (Figure \ref{fig:covariate_shift}), and potentially other variables which show similar degrees of covariate shift. 

Another potential limitation of our approach is that we use an NNGP kernel based on densely connected neural networks, rather than convolutional networks \citep{garriga2018deep} or recurrent networks \citep{alemohammad2020recurrent}, i.e. we do not explicitly account for spatial and temporal correlations through the model structure. Convolutional NNGP kernels could improve our method's ability to synthesize spatial information, while recurrent kernels could help explicitly account for temporal information. Imposing these relational structures has been shown to help with generalization \citep{battaglia2018relational}. We leave the investigation of these kernels to future research.

Finally, the uncertainty quantification abilities of our model are relatively primitive since they only provide a single variance value for each field-wise prediction. Our method tended to be less calibrated (Table \ref{tab:interval_metrics}) than methods with spatially varying uncertainty quantification (EA, WEA, and LM). However, we showed that NN-GPR represented uncertainty better in high-variance regions, such as over land for surface temperatures. Theoretically, our model could be extended to incorporate spatially varying uncertainty, which may improve prediction and uncertainty quantification. However, computationally, this would make likelihood evaluation prohibitively expensive since we could no longer use the Kronecker product shortcut for matrix inversion (Section A).

The simulation results in Sections \ref{sec:point_estimation} and \ref{sec:uncertainty_quantification} showed that NN-GPR is most accurate when the test data are relatively similar to the training data, i.e., in the near future rather than the far future. This suggests that the NN-GPR approach could be quite powerful in weather prediction, where vast quantities of high-frequency training data are available, and predictions only need to be made over short time horizons (a few days). In future work, we will explore the scalability of the NN-GPR approach to high-frequency weather data and evaluate its competitiveness against leading finite-width deep learning approaches. We will also explore methods for including explicit spatial and temporal information in the predictions, and non-Gaussian likelihoods, such as for extremes.

\begin{acks}[Acknowledgments]
The authors would like to thank the anonymous referees, an Associate
Editor and the Editor for their constructive comments that improved the
quality of this paper.
\end{acks}

\begin{funding}
B. Li’s research is partially supported by NSF-DMS-1830312 and NSF-DMS-2124576. R. Sriver was partially supported by the U.S. Department of Energy, Office of Science, Biological and Environmental Research Program, Earth and Environmental Systems Modeling, MultiSector Dynamics, Contracts No. DE-SC0016162 and DE-SC0022141.
\end{funding}


\begin{supplement}
    \sname{Appendix}
    \sfilename{NNGP_supplement.pdf}
    \sdescription{Additional tables, figures, and details are included in the Appendix.}
\end{supplement} 

\begin{supplement} 
    \sname{Code}
    \slink[url]{https://github.com/trevor-harris/nngpr}
    \sdescription{Github repository contains Python code implementing our model and scripts that allow for reproducing paper results.}
\end{supplement}

\bibliographystyle{imsart-nameyear}
\bibliography{nngp}

\appendix
\section{Parameter Estimation} \label{sec:prior_optimization}

We use $\theta = \{\sigma^2_w, \sigma^2_b \}$ to denote the weight and bias variances in the neural network, and $\phi = \{\theta, \sigma^2 \}$ to denote the full parameter collection including the error variance. For simplicity, and without loss of generality, we will assume $X\beta$ is zero. Because we further assume $\phi$ is shared across all spatial locations and all locations are independent, we can write the joint distribution over $Y_i(p)$ for all $i \in 1,...,n$ and $p \in s_1,...,s_d$, as
\begin{equation}\label{eqn:joint_likelihood}
   N(0, \Sigma(\theta)_{n \times n} \otimes I_d + \sigma^2 I_{nd}),
\end{equation}
where $\otimes$ denotes the Kronecker product. To maximize the likelihood, we take the negative log-likelihood, which is proportional to
\begin{equation} \label{eqn:original_loss}
	\mathcal{L}(\phi, D) = \frac{1}{2}Y^T K^{-1}(\phi)Y + \frac{1}{2} \log|K(\phi)|.
\end{equation}

For most practical problems, the dimension of the reanalysis fields, $d$, is too large to compute the true $nd \times nd$ covariance matrix $K(\phi)$, let alone invert $K(\phi)$. However, we can avoid unnecessary computations and make matrix inversion manageable by exploiting the block diagonal structure of the covariance matrix. That is, We can make evaluations of $\mathcal{L}(\phi, D)$ tractable by using the fact that
\begin{align*}
	K(\phi) &= \Sigma(\theta) \otimes I_d + \sigma^2 I_{nd} \\
	          &= (\Sigma(\theta) + \sigma^2 I_{n}) \otimes I_d,
\end{align*}
since we assumed each location has the same noise distribution  $N(0, \sigma^2)$.
Therefore, $K^{-1}(\phi) = (\Sigma(\theta) + \sigma^2 I_{n})^{-1} \otimes I_d^{-1}$ and 
\begin{align*}
	Y^T K^{-1}(\theta)Y = \sum_p  Y^T_p (\Sigma(\theta) + \sigma^2 I_{n})^{-1} Y_p,
\end{align*}
where $Y_p$ is a $1 \times n$ vector consisting of the $p'th$ location in each of the $n$ reanalysis fields. For our model integration problem, $n$ is typically small enough ($n < 1000$) to quickly invert the $n \times n$ matrix $(\Sigma(\theta) + \sigma^2 I_{n})$. Similarly, we use $K(\phi)  =(\Sigma(\theta) + \sigma^2 I_{n}) \otimes I_p$, to get
\begin{align*}
	\log|K(\phi)| &= \log|(\Sigma(\theta) + \sigma^2 I_{n}) \otimes I_p| \\
	       &= p\log|\Sigma(\theta) + \sigma^2 I_{n}| + n \log|I_p| \\
	       &= p\log|\Sigma(\theta) + \sigma^2 I_{n}|.
\end{align*}
These two manipulations result in an equivalent loss function (to equation \ref{eqn:original_loss})
\begin{equation} \label{eqn:efficient_loss}
	\mathcal{L}(\phi, D) = \sum_p  Y^T_p (\Sigma(\theta) + \sigma^2 I_{n})^{-1} Y_p + p\log|\Sigma(\theta) + \sigma^2 I_{n}|,
\end{equation}
with a significantly lower memory cost.

The loss in Equation \ref{eqn:efficient_loss} does not readily admit closed-form solutions for any parameter in $\theta$. However, for some common activation functions, such as ReLU or Tanh, $\mathcal{L}(\theta, D)$ can be differentiated with respect to $\theta$, since $\Sigma(\theta)$ is differentiable with respect to $\theta$. We can, therefore, apply gradient descent or any of its variants to find local minimizers $\hat{\theta}$. Stochastic gradient descent is also possible with mini-batching performed over the $p$ locations rather than the $n$ samples.

Optimizing the prior, rather than setting it, is crucial since, as shown in \citep{lee2017deep}, the two priors determine how well the neural network $\Phi_A(\cdot, \theta)$ can learn. The variance parameters and the activation function define a phase space, with large regions corresponding to poor predictive performance. Manually setting the parameter values is precarious since the critical region where training can occur can be small and shrinks with increasing depth. We optimize their values to locate this critical region to help ensure our model (Equation 7 in the manuscript) has low approximation and generalization error.

\section{Additional Experiments} \label{sec:additional_experiments}

\subsection{Lagged Inputs} \label{sec:lagged_inputs}

Notably, our model (Equation 7 in the manuscript) does not explicitly account for shifts in time, i.e. the influence of past or future ensemble values. However, we can accommodate this type of temporal dependence by concatenating the ensemble at time $t$ with the ensemble at time $t+1$, $t-1$, $t-2$, etc and using the concatenated ensembles as inputs to NN-GPR. Lagged ensembles provide additional information and could potentially improve predictive performance.

    \begin{figure}
    \includegraphics[width=1\textwidth]{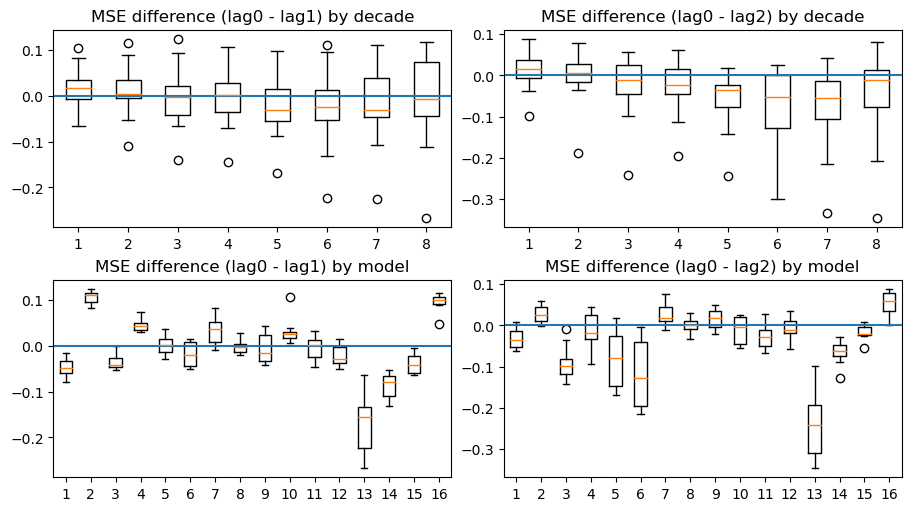}
    \caption{Boxplots of the difference between NNGPR with no lagged ensembles (lag0), a single backward time lag $t-1$ (lag1), and one backward and one forwards lag $t-1$ and $t+1$, respectively (lag2). In the lagged cases the contemporary ensemble at time $t$ is also included.}
    \label{fig:lagged_differences}
    \end{figure}

We conducted a small simulation study to compare NN-GPR's performance with and without lagged models for T2M prediction. We compare three settings:
    \begin{itemize}
        \item[1.] lag0 -- NN-GPR using the ensemble from time $t$ input (standard NN-GPR)
        \item[2.] lag1 -- NN-GPR using the ensembles from time $t$ and $t-1$ as input
        \item[3.] lag2 -- NN-GPR using the ensembles from time $t+1$, $t$, and $t-1$ as input
    \end{itemize}
Using each of these three models, we replicated the T2M portion of the perfect model experiments and summarize the results with MSE (Figure \ref{fig:lagged_differences}). Figure \ref{fig:lagged_differences} shows no MSE improvement by including lagged ensembles either by decade or by model being predicted. The unlagged version (lag0) may even slightly improve over lag2 as we predict farther into the future.

\subsection{Hyperparameter Sensitivity}  \label{sec:hyperparameter_sensitivity}

As mentioned in Section 3, there are several hyperparameters that govern the behavior of the NNGP, such as depth, layer type, and activation function. We conducted a sensitivity analysis to study the effect of network depth on our results. 

\begin{figure}
    \centering
    \includegraphics[width=0.95\textwidth]{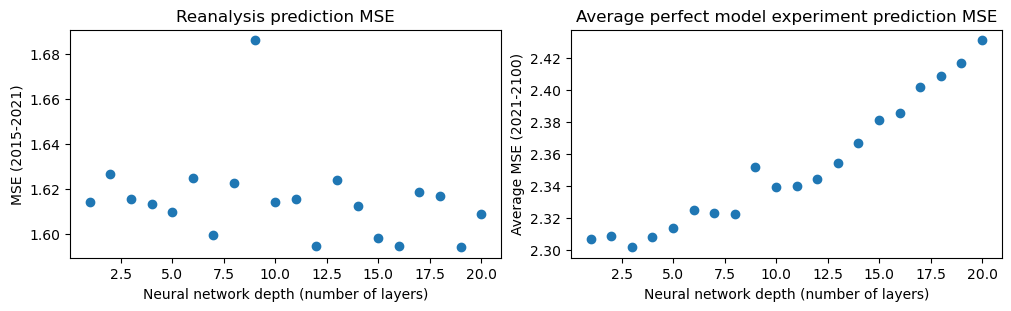}
    \caption{\textbf{Left}: Reanalysis MSE by network depth. \textbf{Right}: Average (over 16 experiments) perfect model MSE by network depth.} \label{fig:sensitivity}
\end{figure}

We repeat the T2M perfect model experiments (Section 4) with 20 different NN-GPR models by varying the depth from 1 to 20. We computed the MSE of NN-GPR's predictions under each depth setting (Figure \ref{fig:sensitivity} right side) and averaged this over the 16 perfect model runs. This allows the effect of depth on forecasting skill over long time scales (roughly 80 years). We then repeat our T2M reanalysis experiments (Section 5), again with 20 different NN-GPR models by varying the depth from 1 to 20. We computed the MSE for each depth setting (Figure \ref{fig:sensitivity} left side). This allows us to asses the effect of depth on short term forecasting skill (6 years) under realistic settings.

In both cases, there is a definite trend (either positive or negative) with layer depth, but the spread between MSE from best to worst mode is not drastically large (Figure \ref{fig:sensitivity}). In the model experiments, we found a clear, though small, positive correlation between depth and MSE. Thus higher depth values resulted in higher MSEs for long term forecasting. Here the optimal setting was depth = 3. In the reanalysis experiments, found a slight negative trend, meaning higher depth values resulted in lower MSEs for short term forecasting. The differences were essentially negligible, but the lowest MSE setting was depth = 12. We also tested short term prediction with perfect model experiments (not shown) and found the same result as in the reanalysis predictions. 

We chose a depth of 10 as a compromise between short term prediction skill, long term prediction skill, and computational effort. In Section B, we compare the performance of NN-GPR using a range of depths from 1 to 20. We find that predictive skill is relatively insensitive to network depth although shallower networks perform slightly better for long term prediction and deep networks perform slightly better short term prediction.

\subsection{Comparison between NN-GPR and GPSE} \label{sec:nngp_gpse_comparison}

Because NN-GPR and GPSE are relatively close in performance in Table 1, we investigate their systematic differences more closely. We investigate the MSE difference between NN-GPR and GPSE \textit{within} each perfect model experiment. That is we want to see, for any given experiment or prediction period, if NN-GPR will have a lower MSE and higher SSIM than GPSE.

 \begin{figure}
    \includegraphics[width=1\textwidth]{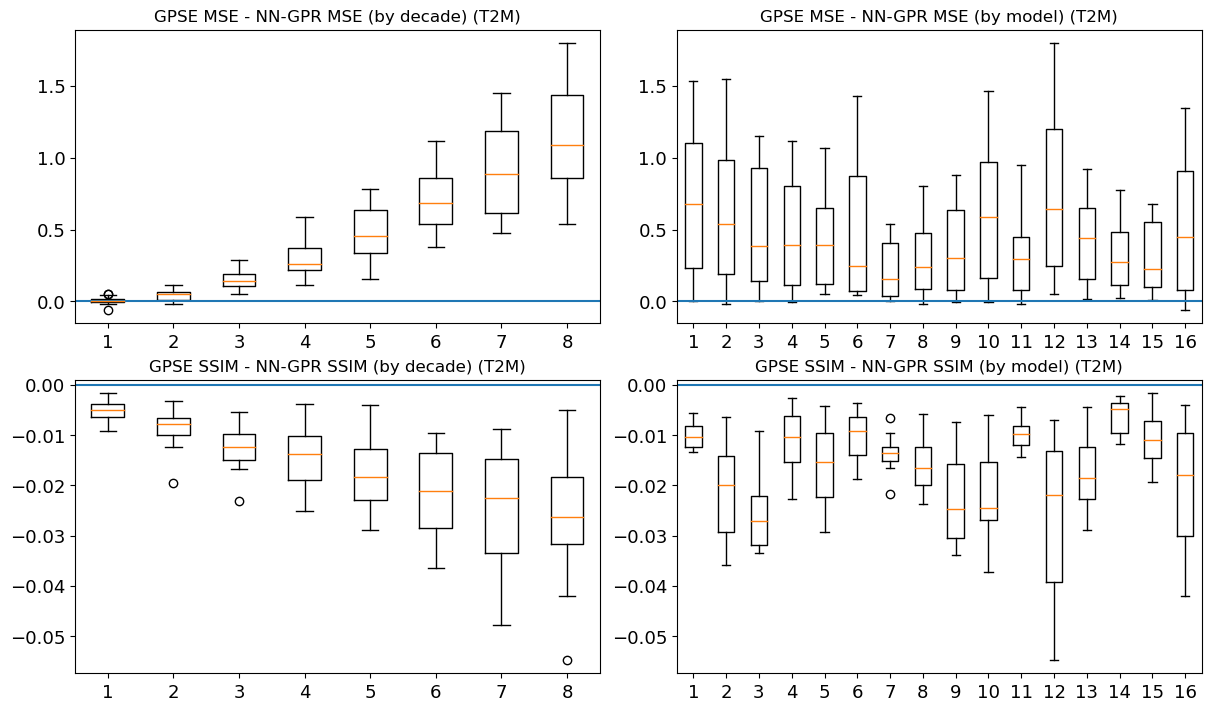}
    \includegraphics[width=1\textwidth]{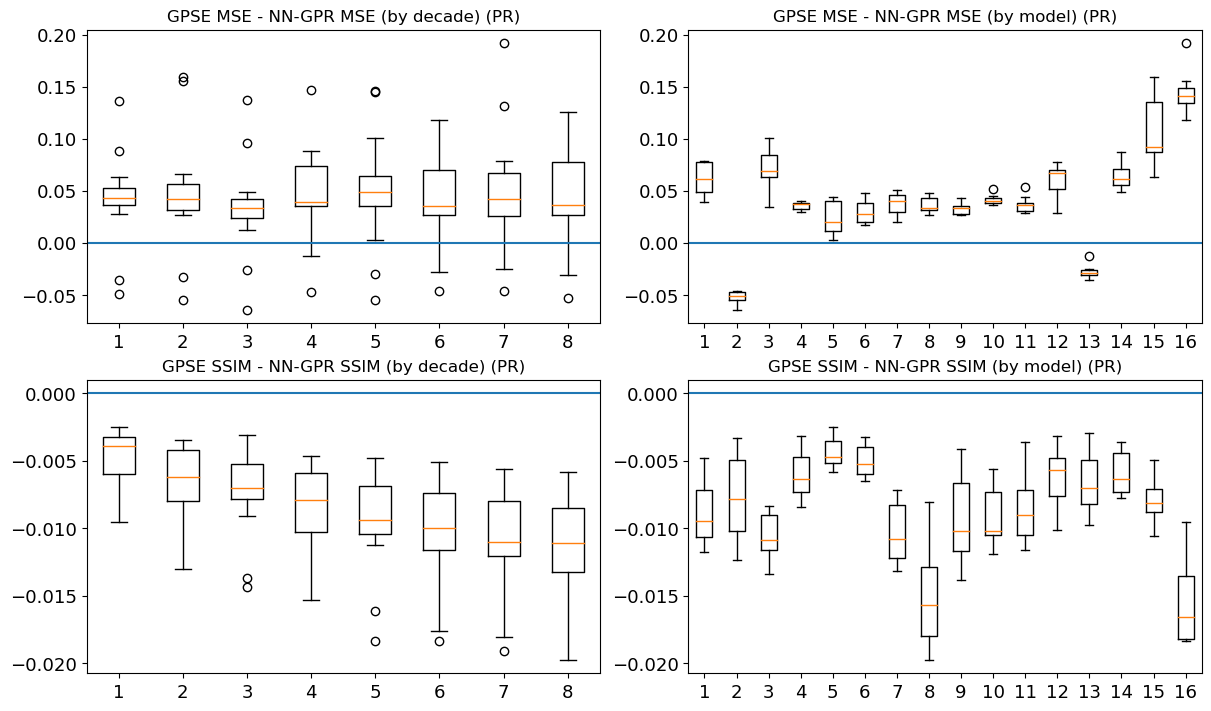}
    \caption{Boxplots of the difference between GPSE and NN-GPR for T2M and PR. Distribution of the differences shows a clear bias. GPSE always has higher MSE in T2M and usually higher MSE in PR. GPSE always has lower (worse) SSIM in T2M and PR.} \label{fig:within_experiment_metrics}
\end{figure}

Figure \ref{fig:within_experiment_metrics} is broken down by prediction decade (1-8), experiment number (1-16), and variable type (T2M and PR). Figure \ref{fig:within_experiment_metrics} shows MSE differences by decade (left column) and by model (right column) for temperature (top two rows) and precipitation (bottom two rows). For temperature it is clear that for almost every decade and every model, GPSE has a higher MSE and lower SSIM as evidenced by the boxplots being completely over (for MSE) or under (for SSIM) the blue 0 line. The only exception is in the first two decades where the MSE differences are comparable. For precipitation, the results are less unanimous, but show that GPSE generally has a higher MSE (by model and year) and a uniformly lower SSIM (by model and year). We conclude that, while the average metrics are close in Table 1, if we look at any given prediction task or prediction time period that NN-GPR shows a small but clear advantage.

\subsection{Miscalibration Error} \label{sec:Miscalibration Error}

\begin{figure}
\centering
\includegraphics[width=0.9\textwidth]{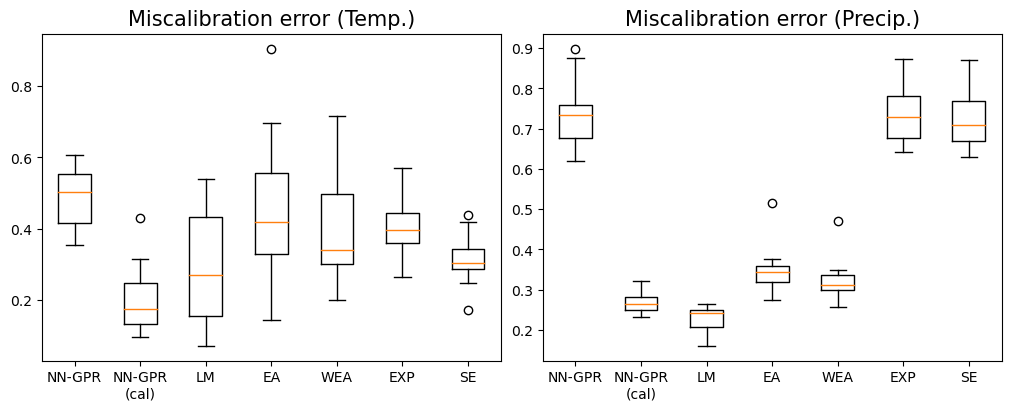}
\caption{Miscalibration error as measured by PIT for the NN-GPR, LM, EA, WEA, GPSE, and GPEX. We also include NN-GPR (cal) which uses the predictions from NN-GPR and the predictive variance from LM. NN-GPR shows the poorest average calibration on temperature predictions, and is comparable with GPSE and GPEX on precipitation. Calibration is measured on reanalysis field predictions from 2015-2021 (same data as Section 5).} \label{fig:pit}
\end{figure}

Figure \ref{fig:pit} shows the miscalibration error ($L_2$ distance between PIT and a uniform density) for each method except the CNN. NN-GPR is the least calibrated for temperature prediction, and comparable with GPSE and GPEX for precipitation. Some miscalibration is likely due to NN-GPR having constant variance leading to simultaneous over and under estimation of the variance in different regions. To test this idea we created NN-GPR (cal) which uses the predictions from NN-GPR and the variance from LM. The calibration of NN-GPR (cal) is comparable to the approaches with spatially varying variance (LM, EA, and WEA). Ideally, we would incorporate spatially varying variance directly in our model but the direct likelihood approach is computationally infeasible (covariance matrix increases from $n \times n$ to $nd \times nd$ for $d$ spatial locations), so we leave this to future efforts. Miscalibration could also be due in part to the ReLU activation leading to overconfidence \cite{kristiadi2020being}, although we do not prove this here.

\bibliographystyle{imsart-nameyear}
\bibliography{nngp}

\end{document}